\documentclass[11pt,letter]{article}

\usepackage[left=2.5cm,right=2.5cm,top=2.5cm,bottom=2.5cm,includefoot]{geometry}
\usepackage{xcolor}
\usepackage[small]{titlesec}
\titlelabel{\thetitle.\quad}
\usepackage{endnotes}
\usepackage{setspace}
\usepackage{placeins}
\usepackage{grffile} 
\usepackage{footmisc}
\DefineFNsymbols{mySymbols}{{\ensuremath\dagger}{\ensuremath\ddagger}\S\P
   *{**}{\ensuremath{\dagger\dagger}}{\ensuremath{\ddagger\ddagger}}}
\setfnsymbol{mySymbols}
\usepackage[title]{appendix}

\usepackage[english]{babel}
\usepackage[utf8]{inputenc}
\usepackage[T1]{fontenc}
\usepackage[protrusion=true,expansion=true]{microtype}
\usepackage{newtxtext}
\usepackage{amsthm}  
\usepackage[smallerops]{newtxmath}
\usepackage{dsfont}
\numberwithin{equation}{section}
\DeclareMathAlphabet{\mathcal}{OMS}{cmsy}{m}{n}

\usepackage{stackengine}
\usepackage{booktabs}
\usepackage{enumitem}
\setlist[enumerate]{align=left}
\usepackage{listings}
\usepackage{multirow,multicol}
\usepackage[font={footnotesize}]{caption}
\usepackage[font={footnotesize}]{subcaption}
\usepackage{tabularx}
\usepackage{rotating} 
\usepackage{wrapfig}

\newcommand*{\thead}[1]{\multicolumn{1}{c}{\begin{tabular}{@{}r@{}}#1\end{tabular}}}
\newcommand*{\theadl}[1]{\multicolumn{1}{l}{\begin{tabular}{@{}l@{}}#1\end{tabular}}}

\usepackage{colortbl}
\usepackage{siunitx}
\usepackage{diagbox}
\newtheoremstyle{boldtitle}
{}
{}
{}
{}
{\bfseries}
{.}
{.5em}
{{\thmname{#1 }}{\thmnumber{#2}}{\thmnote{ (#3)}}}

\theoremstyle{boldtitle}

\newtheorem{theorem}{Theorem}[section]
\newtheorem{lemma}[theorem]{Lemma}
\newtheorem{definition}[theorem]{Definition}

\newtheorem{corollary}[theorem]{Corollary}

\newtheorem{assumption}[theorem]{Assumption}

\usepackage{xr-hyper} 
\usepackage[
    breaklinks,
    colorlinks=true,
    pagebackref=false,
    pdfauthor={Sebastian Bayer and Timo Dimitriadis},%
    pdftitle={Regression Based Expected Shortfall Backtesting},%
    citecolor=blue,
    linkcolor=blue,
    urlcolor=blue
]{hyperref}      
\usepackage{cleveref}
\usepackage[authoryear,round]{natbib} 
\DeclareMathOperator{\ES}{ES}

\newcommand\ubar[1]{\stackunder[1.2pt]{$#1$}{\rule{.8ex}{.075ex}}}

\DeclareMathOperator*{\argmin}{arg\,min}


\begin{document}

\title{Regression Based Expected Shortfall Backtesting}

\author{Timo Dimitriadis\thanks{Corresponding Author, Heidelberg Institute for Theoretical Studies, Heidelberg, Germany and University of Hohenheim, Germany, e-mail: \texttt{timo.dimitriadis@h-its.org} } 
	\and Sebastian Bayer\thanks{University of Konstanz, Konstanz, Germany, e-mail: \texttt{sebastian.bayer@uni-konstanz.de}} 
}

\maketitle

\vspace*{1em}

\noindent\rule{\linewidth}{.4pt}
\textbf{Abstract}
\\[.5\baselineskip]
\noindent
This paper introduces novel backtests for the risk measure Expected Shortfall (ES) following the testing idea of \cite{MincerZarnowitz1969}.
Estimating a regression framework for the ES stand-alone is infeasible, and thus, our tests are based on a joint regression for the Value at Risk and the ES, which allows for different test specifications.
These ES backtests are the first which solely backtest the ES in the sense that they only require ES forecasts as input parameters.
As the tests are potentially subject to model misspecification, we provide asymptotic theory under misspecification for the underlying joint regression.
We find that employing a misspecification robust covariance estimator substantially improves the tests' performance.
We compare our backtests to existing approaches and find that our tests outperform the competitors throughout all considered simulations.
In an empirical illustration, we apply our backtests to ES forecasts for 200 stocks of the S\&P\,500 index.
\\[1em]
\\[-.5\baselineskip]
\noindent\rule{\linewidth}{.4pt}


\section{Introduction}

Through the transition from Value at Risk (VaR) to Expected Shortfall (ES) as the primary market risk measure in the Basel Accords \citep{Basel2016, Basel2017}, there is a great demand for reliable methods for estimating, forecasting and backtesting the ES.
Formally, the ES at level $\tau \in (0,1)$ is defined as the mean of the returns smaller than the respective $\tau$-quantile (the VaR), where $\tau$ is usually chosen to be 2.5\% as stipulated by the Basel Accords.
The ES is introduced into the banking regulation because it overcomes several shortcomings of the VaR, such as being not coherent and its inability to capture tail risks beyond the $\tau$-quantile \citep{Artzner1999,Danielsson2001,Basel2013}. 
In contrast to estimation and forecasting of ES where most of the existing models for the VaR can easily be adapted and generalized to the ES, such a generalization is not as straight-forward for backtesting ES forecasts \citep{Emmer2015}.
In general, backtesting of a risk measure is the process of testing whether given forecasts for this risk measure are correctly specified, which is carried out by comparing the history of the issued risk forecasts with the corresponding realized returns.
The primary difficulty in directly backtesting ES is its non-elicitability and non-identifiability \citep{Weber2006, Gneiting2011, Fissler2016, Fissler2016b} as consequently, there is no analog to the hit sequence which is the natural identification function of quantiles and which lies at the heart of almost all VaR backtests.%
\footnote{See \cite{Yamai2002, Kerkhof2004, Carver2013, Acerbi2014, Emmer2015, Ziegel2016, Fissler2016b, Nolde2017} for the ongoing discussion on backtestability of the ES.}

As a consequence, most of the proposed procedures in the growing literature on backtesting ES use indirect approaches by formally backtesting some quantity which is closely related to the ES.
Examples include tests based on the entire tail distribution, a linear approximation of the ES through several quantiles or the pair consisting of the VaR and the ES.\footnote{In particular, several tests require the whole or tail distribution of the returns or equivalently the cumulative violation process \citep{Kerkhof2004, Wong2008, Graham2014, Acerbi2014, Du2017, Loeser2018, CostanzinoCurran2018}, multiple quantiles at different levels \citep{Emmer2015, Costanzino2015, Kratz2018, CouperierLeymarie2019}, the VaR and the volatility \citep{McNeil2000,Nolde2017,Righi2013,Righi2015}, or the VaR \citep{McNeil2000,Nolde2017} in addition to the ES forecasts.
See Appendix \ref{sec:existing_backtests} for an overview over the existing backtesting approaches.}
We argue that formally, these approaches are backtests for the auxiliary quantities rather than for the ES itself, see also \cite{Nolde2017}.
This distinction is particularly important as these backtests require further input parameters such as forecasts for the VaR at multiple levels, the tail distribution beyond some quantile, or even the entire distribution.
The regulatory authorities however do not have this additional information at hand as it is not mandatorily reported by the financial institutions \citep{Aramonte2011,Basel2016,Basel2017}.
As a consequence, the existing, so-called ES backtests are not applicable where they are most needed.


In this paper, we propose novel backtests for ES forecasts which are the first strict ES backtests in the literature in the sense that besides the realized returns, they only require ES forecasts as input parameters.
Our tests  follow the general regression based testing idea of \cite{MincerZarnowitz1969}.
For this, we estimate a regression framework which models the conditional ES at level $\tau$ as a linear function $\ES_{\tau} \left( Y_t  \mid \mathcal{F}_{t-1} \right) = \gamma_1 + \gamma_2 \hat e_{t}$, where we use financial returns $Y_t$ as the response variable and the given ES forecasts $\hat e_{t}$ as the explanatory variable including an intercept term.
For correctly specified ES forecasts, the intercept and slope parameters equal zero and one, which we test for by using a Wald statistic.
As the ES is not elicitable \citep{Gneiting2011}, we face the methodological difficulty that we cannot estimate such a regression framework for the ES stand-alone as neither loss nor identification functions are available for the ES which could be used as objective functions for M- or GMM-estimation \citep{DimiBayer2019}.
Recently, \cite{Patton2019} and \cite{DimiBayer2019} propose a feasible alternative by specifying an auxiliary quantile regression equation $Q_{\tau} \left( Y_t  \mid \mathcal{F}_{t-1} \right) = \beta_1 + \beta_2 \hat \rho_{t}$ (with explanatory variable $\hat \rho_t$) and by jointly estimating the regression parameters $(\beta, \gamma)$ by employing a joint loss function for the quantile and the ES from \cite{Fissler2016}.

The specification of the quantile equation allows for different testing approaches.
First, we employ auxiliary VaR forecasts $\hat v_{t}$ as the explanatory variable in the quantile equation, but only test the ES specific parameters $\gamma$. 
We refer to this test as the \textit{Auxiliary ESR} (ES Regression) backtest.
The main drawback of this test is that it requires auxiliary VaR forecasts and consequently, it is formally a joint backtest for the VaR and ES which, however, mainly focuses on the ES by only testing the ES specific regression parameters.
Second, we use the ES forecasts $\hat e_{t}$ as the explanatory variable in both, the quantile and the ES equation and again only test on the ES specific parameters $\gamma$. 
We refer to this test as the \textit{Strict ESR} backtest as it only requires ES forecasts as input parameters and consequently is the first test in the literature which solely backtests ES forecasts.
This testing idea comes at the drawback of a potential model misspecification in the quantile equation if the underlying data goes beyond a pure scale (volatility) model.
Therefore, we provide asymptotic theory for this joint quantile and ES regression framework under model misspecification, which generalizes the asymptotic theory introduced in \cite{DimiBayer2019} and \cite{Patton2019}. 
The potential model misspecification results in a more complex and usually inflated asymptotic covariance matrix.
We account for this in the implementation of our tests by employing a new covariance estimation technique which explicitly estimates these new covariance terms.

We further introduce an intercept variant of the Strict ESR backtest by fixing the slope parameter in the regression to one, and by only estimating and testing the intercept term.
We refer to this backtest as the \textit{Intercept ESR} backtest.
This test allows for both, testing against one-sided and two-sided alternatives.
In contrast, the other two proposed ESR backtests only allow for testing against two-sided alternatives as it is generally unclear how underestimated and overestimated ES forecasts influence the intercept and slope  parameters.
Because the capital requirements that the financial institutions must keep as a reserve depend on the reported risk forecasts, the market participants have an incentive to report risk forecasts which are \textit{too risky} in order to minimize the expensive capital requirements.
In contrast, issuing too conservative risk forecasts results in larger capital reserves, which does not have to be punished by the regulatory authorities.
Thus, the regulators only have to prevent and penalize the underestimation of the financial risks, which demonstrates the necessity of one-sided testing procedures.
For example, the currently applied traffic light system \citep{Basel1996} is in fact a one-sided VaR backtest.
As the Strict ESR backtest, the Intercept ESR backtest also has the desired characteristic to only require ES forecasts as input parameters and consequently is the first procedure that solely backtests the ES against a one-sided alternative.
We provide implementations of the three ESR backtests proposed in this paper in the R package \texttt{esback} \citep{Bayer2019a}.

Such regression-based forecast evaluation approaches are already used for testing mean forecasts \citep{MincerZarnowitz1969}, quantile forecasts \citep{Gaglianone2011,Guler2017}, and expectile forecasts \citep{Guler2017}.
In contrast to these functionals, where regression techniques are easily available (see e.g. \citealp{Koenker1978}, \citealp{Efron1991}), the non-elicitability of the ES makes our approach more involved but also opens up the possibility for the different testing specifications we introduce.
Our multivariate generalization approach of the \cite{MincerZarnowitz1969} testing idea can be applied equivalently to other \textit{higher-order elicitable} functionals \citep{Fissler2016} such as e.g. the variance (in the presence of a non-zero mean) and the Range VaR \citep{Cont2010, Embrechts2018}.

We evaluate the empirical properties of our ESR backtests and compare them to the existing joint VaR and ES backtests of \citet{McNeil2000} and \citet{Nolde2017} through several simulation designs.
In the first setup, we implement the classical size and power analysis for backtesting risk measures, where we simulate data stemming from several realistic data generating processes and evaluate the empirical rejection frequencies of the backtests for forecasts stemming from the true and from some misspecified forecasting model.
In order to assess how the potential model misspecification affects the Strict and the Intercept ESR backtests, we utilize DGPs which go beyond the class of pure scale (volatility) processes.
For this, we implement two different Student's-$t$ GAS models with time-varying higher moments \citep{Creal2013} and furthermore use an AR-GARCH model which allows for gradually increasing the degree of misspecification through the AR parameter.
In the second setup, we introduce a new technique for evaluating the power of backtests for financial risk measures, where we continuously misspecify certain model parameters of the data generating process to obtain a continuum of alternative models with a gradually increasing degree of misspecification.
Misspecifying the different model parameters separately allows us to misspecify certain model characteristics (such as the reaction to shocks) in isolation, which permits a closer examination of the proposed backtesting procedures.

The simulations show that all three ESR backtests we propose in this paper are well-sized, especially when the tests are applied using the new covariance estimation method which accounts for possible model misspecification.
We further find that the performance of our testing procedures is almost unaffected by the DGPs which cause model misspecification in the Strict and the Intercept ESR tests.
Moreover, our tests are more powerful than the existing backtests of \citet{McNeil2000} and \citet{Nolde2017} in almost all of the considered simulation designs for both, testing against one-sided and two-sided alternatives.
Notably, throughout all simulation designs, the ESR backtests are able to detect the various different misspecifications of the forecasts.
In contrast, the existing backtests sometimes completely fail to detect certain misspecifications, for instance when the forecaster reports risk forecasts for a misspecified probability level.


The rest of this paper is organized as follows.
\Cref{sec:theory} introduces our new ESR backtests and presents asymptotic theory under model misspecification.
\Cref{sec:monte_carlo} contains several simulation studies and \Cref{sec:empirical_application} applies the backtests to ES forecasts for a large amount of stocks from the S\&P\,500 index.
\Cref{sec:conclusion} concludes.
The proofs are deferred to Appendix \ref{sec:Proofs} and Appendix \ref{sec:TechnicalProofs}.

%

\section{Theory}
\label{sec:theory}

\subsection{Setup and Notation}

We consider a stochastic process 
\begin{align}
	\label{eqn:DefinitionStochasticProcess}
	{Z} = \bigl\{ {Z}_t: \Omega \to \mathbb{R}^{l+1}, \, l \in \mathbb{N},\, t = 1,\dots, T \bigr\},
\end{align}
defined on some complete probability space $\bigl( \Omega,\, \mathcal{F},\, \mathbb{P} \bigr)$, with the filtration $\mathcal{F} = \bigl\{ \mathcal{F}_t,\, t = 1,\, \dots, T \bigr\}$ and $\mathcal{F}_t = \sigma \{{Z}_s,\, s\le t \}$ for all $t = 1,\dots,T$, where $T \in \mathbb{N}$.
We partition the stochastic process ${Z}_t = (Y_t, U_t)$, where $Y_t$ is an absolutely continuous random variable of interest and $U_t$ is an $l$-dimensional vector of explanatory variables.
We denote the conditional cumulative distribution function of $Y_t$ given the past information $\mathcal{F}_{t-1}$ by $F_t(y) = \mathbb{P} ( Y_t \le y \mid \mathcal{F}_{t-1})$ and the corresponding probability density function by $f_t$. 
Whenever they exist, the mean and the variance of $F_t$ are denoted by $\mathbb{E}_t[\cdot]$ and $\operatorname{Var}_t(\cdot)$.

For financial applications, the variable $Y_t$ denotes the daily log returns of a financial asset (for instance, a stock or a portfolio), i.e. $Y_t = \log P_t - \log P_{t-1}$, where $P_t$ denotes the price of the asset at day $t = 1,\ldots,T$.
This means that throughout this paper, we use the sign convention that positive returns denote profits, and negative returns denote losses.
The vector $U_t$ contains further variables that are used to produce forecasts for certain functionals (usually risk measures) of the random variable $Y_t$.
We are interested in testing whether forecasts for a certain $d$-dimensional, $d \in \mathbb{N}$ functional (risk measure) $\rho = \rho(F_t)$ of the conditional distribution $F_t$ are correctly specified.
For that, we define the most frequently used functionals for financial risk management in the following.
The conditional quantile of $Y_t$ given the information set $\mathcal{F}_{t-1}$ at level $\tau \in (0,1)$ is defined as
$Q_\tau \bigl(Y_t \mid \mathcal{F}_{t-1} \bigr) = F_t^{-1}(\tau) = \inf \bigl\{ y \in \mathbb{R}: F_{t}(y) \ge \tau \bigr\}$,
which is called the VaR at level $\tau$ in financial applications.
Furthermore, we define the functional ES at level $\tau$ of $Y_t$ given $\mathcal{F}_{t-1}$ as
$\ES_\tau \bigl( Y_t \mid \mathcal{F}_{t-1} \bigr) = \frac{1}{\tau} \int_0^\tau F_t^{-1}(s) \, \mathrm{d}s$.
If the distribution function $F_t$ is continuous at its $\tau$-quantile, this definition can be simplified to the truncated tail mean of $Y_t$,
\begin{align}
	\ES_\tau \bigl( Y_t \mid \mathcal{F}_{t-1} \bigr) = \mathbb{E}_t \left[ Y_t \mid  Y_t \leq Q_\tau \bigl( Y_t \mid \mathcal{F}_{t-1} \bigr) \right].
\end{align}
We denote an $\mathcal{F}_{t-1}$-measurable one-step-ahead forecast for day $t$ for the risk measure $\rho$ of the distribution $F_t$, stemming from some external forecaster or from some given forecasting model%
\footnote{For recent overviews on VaR and ES forecasting approaches, see \citet{Komunjer2013} and \citet{Nadarajah2014}.}
by $\hat \rho_t = \hat \rho_t(\mathcal{F}_{t-1})$.
Following this notation, we denote forecasts for the $\tau$-VaR by $\hat v_t$ and for the $\tau$-ES by $\hat e_t$ for some fixed level $\tau \in (0,1)$.
For simplicity of the notation, we drop the dependence on $\tau$ as it is a fixed quantity.

As both, the incentive of the forecaster and the underlying method used to generate the forecasts are in general unknown, these forecasts are not necessarily correctly specified. 
The focus of this paper is to develop statistical tests for correctness of a given series of forecasts $\bigl\{ \hat \rho_t,\, t=1,\dots,T \bigr\}$ for the risk measure $\rho$ relative to the realized return series $\bigl\{ Y_t,\, t=1,\dots,T \bigr\}$.
This is  in the literature usually referred to as \textit{backtesting} of the risk measure $\rho$ without strictly defining this terminology.
We provide such a definition in the following.

\begin{definition}
	\label{def::ProperBacktest}
	A \textit{backtest} for the series of forecasts $\bigl\{ \hat \rho_t,\, t=1,\dots,T \bigr\}$ for the $d$-dimensional risk measure (functional) $\rho$ relative to the realized return series $\bigl\{ Y_t,\, t=1,\dots,T \bigr\}$ is a function
	\begin{align}
		f: \mathbb{R}^T \times \mathbb{R}^{T \times d} &\to \{ 0, 1\},
	\end{align}
	which maps the return and forecast series onto the respective test decision.
\end{definition}

The core message of this definition is that besides the realized return series, a backtest for some risk measure is only allowed to require forecasts for this risk measure as input parameters.
This strict differentiation becomes relevant in the context of backtesting ES as, in contrast to the existing VaR backtests, the recently proposed ES backtests require further input parameters such as forecasts for the VaR, the volatility, or the entire tail distribution.
The demand for these further quantities induces the following practical problems.
First, the regulatory authorities who rely on such backtesting methods do not necessarily receive forecasts from the financial institutions for the additional information required by these tests, which makes such backtests inapplicable for the regulatory authorities.
Second, a rejection of the tests does not necessarily imply that the ES is misspecified, but that the forecasts for any of the input components are misspecified.
Consequently, these tests are in fact not backtests for the ES, but rather backtests for some vector of risk measures (or the entire tail distribution).


\subsection{The ESR Backtests}
\label{sec::mz_backtest}

We propose backtests for the risk measure ES that test whether a series of ES forecasts $\{\hat e_t, t = 1,\dots T \}$, stemming from some external forecaster or forecasting model, is correctly specified relative to a series of realized returns $\{Y_t, t = 1,\dots, T\}$.
We follow the general testing idea of \cite{MincerZarnowitz1969} and regress the returns $Y_t$ on the forecasts $\hat e_{t}$ and an intercept term by using a regression equation designed specifically for the functional ES,
\begin{align}
    \label{eqn::BacktestRegressionEquation}
    Y_t = \gamma_1 + \gamma_2 \hat e_{t} + u^e_t,
\end{align}
where $\ES_{\tau}(u^e_t  \mid \mathcal{F}_{t-1}) = 0$ almost surely.
Given the structure in (\ref{eqn::BacktestRegressionEquation}) and since the forecasts $\hat e_t$ are generated by using the information set $\mathcal{F}_{t-1}$, this condition on the error term is equivalent to
\begin{align}
	\label{eqn::BacktestFunctionalEquation}
    \ES_{\tau} \left( Y_t  \mid \mathcal{F}_{t-1} \right) = \gamma_1 + \gamma_2 \hat e_{t}.
\end{align}
We then test the hypothesis
\begin{align}
	\label{eqn::NullHypothethis}
	\mathbb{H}_0: ( \gamma_1, \gamma_2 ) = (0,1) \qquad \text{against} \qquad \mathbb{H}_1: (\gamma_1, \gamma_2 ) \neq (0,1).
\end{align}	
Under $\mathbb{H}_0$, the ES forecasts are correctly specified as it holds that $\hat e_t = \ES_{\tau} \left( Y_t \mid \mathcal{F}_{t-1} \right)$ almost surely.%
\footnote{
	Given that the ES forecasts are correctly specified, i.e. $\hat e_t = \ES_{\tau} \left( Y_t \mid \mathcal{F}_{t-1} \right)$, the correct specification condition (\ref{eqn::BacktestFunctionalEquation}) is equivalent to $\gamma_1 = (1-\gamma_2) \hat e_t$.
	This results in the remark of \cite{Holden1990}, who claim that the null hypothesis, given in (\ref{eqn::NullHypothethis}) is only a sufficient, but not a necessary condition for correctly specified forecasts as $\gamma_1 = (1-\gamma_2) \hat e_t$ is the required necessary condition.
	However, this more general condition implies that the forecasts $\hat e_t$ are constant for all $t = 1,\dots,T$, which is highly unrealistic given the dynamic nature of financial time series.
	Consequently, we employ the hypotheses given in (\ref{eqn::NullHypothethis}) for our backtesting procedure.
}
In general, (\ref{eqn::BacktestRegressionEquation}) is an example of a linear regression equation for the ES of the form
	$Y_t = W_t^\top \gamma+ u^e_t$,
for some general vector of covariates $W_t$.
As outlined in \citet{DimiBayer2019} and \cite{Patton2019}, estimating the parameters $\gamma$ 
by M- or GMM-estimation stand-alone is not possible since there do not exist strictly consistent loss and identification functions for the functional ES \citep{Gneiting2011}.
Based on the seminal work of \cite{Fissler2016} who introduce joint loss and identification functions for the VaR and ES, \citet{DimiBayer2019}, \cite{Patton2019} and \cite{Barendse2017} propose the joint regression technique,
\begin{align}
	Y_t = V_t^\top \beta + u^q_t, \label{eqn:JointVaRESreg} \qquad \text{ and } \qquad 
	Y_t = W_t^\top \gamma + u^e_t, 
\end{align}
where $V_t$ and $W_t$ are $k$-dimensional, $\mathcal{F}_{t-1}$-measureable covariate vectors and where 
$Q_{\tau}(u^q_t  \mid \mathcal{F}_{t-1} ) = 0$ and  $\ES_{\tau}(u^e_t  \mid \mathcal{F}_{t-1} ) = 0$ almost surely.
Setting up this joint regression framework facilitates the estimation of the joint regression parameters $(\beta, \gamma)$, whereas stand-alone estimation of $\gamma$ is infeasible.
We use this joint regression setup to propose the following regression based backtests for the ES:
\begin{enumerate}
	\item[\textbf{The Auxiliary ESR Backtest}]
	We choose $V_t = (1, \hat v_t)$ and  $W_t = (1, \hat e_t)$, i.e. we set up the regression system
	\begin{align}
	\label{eqn:RegSpecVaRAndES}
	Y_t = \beta_1 + \beta_2 \hat v_{t} + u^q_t, \qquad \text{ and } \qquad 
	Y_t = \gamma_1 + \gamma_2 \hat e_{t} + u^e_t, 
	\end{align}
	and test
	\begin{align}
	\label{eqn::NullHypothethisAux}
	\mathbb{H}_0: ( \gamma_1, \gamma_2 ) = (0,1) \qquad \text{against} \qquad \mathbb{H}_1:( \gamma_1, \gamma_2 ) \neq (0,1),
	\end{align}	
	using the Wald-type test statistic
	\begin{align}
	\label{eqn:wald_statistic_AuxESR}
	T_{\text{A-ESR}} = T \big( \hat \gamma_T - (0,1) \big) \, \widehat \Omega_{\gamma}^{-1}  \,  \big( \hat \gamma_T - (0,1) \big)^\top,
	\end{align}
	based on some (consistent) covariance estimator $\widehat \Omega_{\gamma}$ for the covariance of the subvector $\gamma$.

	\item[\textbf{The Strict ESR Backtest}]
	We choose  $V_t = W_t = (1, \hat e_t)$, i.e. we set up the regression system
	\begin{align}
		Y_t = \beta_1 + \beta_2 \hat e_{t} + u^q_t, \qquad \text{ and } \qquad 
		Y_t = \gamma_1 + \gamma_2 \hat e_{t} + u^e_t, \label{eqn:RegSpecStrictES}
	\end{align}
	and test
	\begin{align}
		\label{eqn::NullHypothethisStrict}
		\mathbb{H}_0: ( \gamma_1, \gamma_2 ) = (0,1) \qquad \text{against} \qquad \mathbb{H}_1:( \gamma_1, \gamma_2 ) \neq (0,1),
	\end{align}	
	using the Wald-type test statistic
	\begin{align}
		\label{eqn:wald_statistic_StrictESR}
		T_{\text{S-ESR}} = T \big( \hat \gamma_T - (0,1) \big) \, \widehat \Omega_{\gamma}^{-1}  \,  \big( \hat \gamma_T - (0,1) \big)^\top,
	\end{align}
	based on some (consistent) covariance estimator $\widehat \Omega_{\gamma}$ for the covariance of the subvector $\gamma$.
	
\end{enumerate}
We discuss the employed covariance estimators $\widehat \Omega_\gamma$ in Section \ref{sec:TestImplementation}.
Whereas setting up Mincer-Zarnowitz tests for classical elicitable functionals such as the mean, quantiles and expectiles is straight-forward (see \cite{MincerZarnowitz1969}, \cite{Gaglianone2011}, \cite{Guler2017}), in the case of higher-order elicitable functionals such as the ES we have several choices as illustrated above.
The Auxiliary ESR backtest is based on the regression specification (\ref{eqn:RegSpecVaRAndES}) and requires both, VaR and ES forecasts as input parameters.
Thus, following Definition \ref{def::ProperBacktest}, this backtest is formally a joint VaR and ES backtest, however, with a strong emphasis on backtesting ES forecasts.
In contrast, the Strict ESR backtest only incorporates ES forecasts and consequently is the first backtest
for the ES stand-alone.


The Strict ESR test however comes at the cost of a potential model misspecification.
Given that the financial returns $Y_t$ follow some pure scale (volatility) process, it holds that the VaR and ES forecasts are perfectly colinear, $\hat e_t = c \hat v_t$ for some $ c \in \mathbb{R}$.
Consequently, if $\hat v_t$ equals the true conditional VaR, the first equation in (\ref{eqn:RegSpecStrictES}) is correctly specified for the true parameter values $( \beta_1 , \beta_2 ) = (0,c)$.
Most of the financial econometrics literature (almost the entire GARCH, stochastic volatility and Realized Volatility literature) is based on such an assumption for daily returns, which motivates the applicability of this Strict ESR backtest.
However, this backtest is also applicable in the general case where the true VaR and ES forecasts are not necessarily colinear.
For this, we provide asymptotic theory for M-estimation of the joint VaR and ES regression under potential model misspecification in Section \ref{sec:TheoryMisspecification}.

\subsection{The One-Sided Intercept ESR Backtest}
\label{sec:OneSidedTest}

The two ESR backtests introduced in the previous section only allow for testing two-sided hypotheses as specified in (\ref{eqn::NullHypothethisAux})  and (\ref{eqn::NullHypothethisStrict}), as it is generally unclear how too risky (or too conservative) forecasts influence the parameters $\gamma_1$ and $\gamma_2$.
Because the capital requirements the financial institutions have to keep as a reserve depend on the reported risk forecasts, the market participants have an incentive to report too risky forecasts for the ES in order to keep as little capital requirements as possible.
In contrast, issuing too conservative risk forecasts and facing higher capital requirements does not have to be punished by the regulatory authorities.\footnote{One could interpret the higher capital requirements as a punishment for too conservative risk forecasts.}
Thus, the regulators only have to prevent and consequently penalize the underestimation of financial risks, which can be done by using one-sided backtesting procedures.
For example, the traffic light system \citep{Basel1996}, currently implemented in the Basel Accords, is in fact a one-sided backtest for the hit ratios of VaR forecasts.
Consequently, we also introduce a regression-based backtesting procedure for the ES that allows for testing one-sided hypotheses.

\begin{enumerate}
	\item[\textbf{The Intercept ESR Backtest}]	
	This backtest is based on the regression setup of the Strict ESR backtest by regressing the forecast errors, $Y_t - \hat e_t$, on an intercept term only,
	\begin{align}
		\label{eqn::RegressionEquationInterceptTest}
		Y_t - \hat e_t = \beta_1 + u^q_t, \qquad \text{ and } \qquad 
		Y_t - \hat e_t = \gamma_1 + u^e_t,
	\end{align}
	where $Q_\tau (u^q_t\mid\mathcal{F}_{t-1}) = 0$  and $\ES_\tau (u^e_t\mid\mathcal{F}_{t-1}) = 0$ almost surely.
	By using this restricted regression equation, we can define a one-sided and a two-sided alternative,
	\begin{align}
		\begin{split}
			\mathbb{H}_0^{2s}:  \gamma_1 = 0 \qquad    & \text{against} \qquad \mathbb{H}_1^{2s}: \gamma_1 \neq 0, \quad \text{and} \\
			\mathbb{H}_0^{1s}:  \gamma_1 \geq 0 \qquad & \text{against} \qquad \mathbb{H}_1^{1s}: \gamma_1 < 0,
		\end{split}
	\end{align}
	which we test by using a $t$-test based on the estimated asymptotic covariance described in Section \ref{sec:TestImplementation}.
\end{enumerate}	
Note that this testing procedure is equivalent to fixing the slope parameter of the Strict ESR test given in (\ref{eqn:RegSpecStrictES}) to one and only estimating and testing the intercept term.
Therefore, we call this backtest the \textit{Intercept ESR} backtest.

\subsection{Asymptotic Theory under Model Misspecification}
\label{sec:TheoryMisspecification}

In this section, we consider the asymptotic properties of the M-estimator of the joint VaR and ES regression framework given in (\ref{eqn:JointVaRESreg}) under potential model misspecification.
In the following, we write $X_t = (V_t, W_t)$ for the compound vector of covariates.
Following \cite{DimiBayer2019} and \cite{Patton2019}, the M-estimator of the regression parameters $\theta$ is defined by
\begin{align}
	\hat{\theta}_T &= \argmin_{\theta \in \Theta} Q_T(\theta), \qquad \text{ where } \\
	Q_T(\theta) &=  \frac{1}{T} \sum_{t=1}^{T} \rho(Y_t,X_t,\theta) \qquad \text{ and }     \label{eqn:QTObjectiveFunction}      \\
	\rho(Y_t,X_t,\theta) &= \frac{1}{-W_t^\top \gamma} \left( W_t^\top \gamma - V_t^\top \beta + \frac{(V_t^\top \beta - Y_t) \mathds{1}_{\{Y_t \le V_t^\top \beta\}}}{\tau}  \right) + \log(-W_t^\top \gamma), 	\label{eqn::RegressionLossFunction}  
\end{align}
where the loss function in (\ref{eqn::RegressionLossFunction}) is a strictly consistent loss function for the pair quantile and ES \citep{Fissler2016}. 
\cite{DimiBayer2019} and \cite{Patton2019} show consistency and asymptotic normality for the M-estimator in the case of a correctly specified parametric model, i.e under the assumption that there exists a \textit{true} parameter $\theta_0 \in \Theta$ such that
$Q_\tau(u^q_t \mid \mathcal{F}_{t-1}) = 0$ and $ES_\tau(u^e_t \mid \mathcal{F}_{t-1}) = 0$ almost surely.
In the following, we extend this theory by relaxing these assumptions which allows for the general case of misspecified models.
For this, we define the pseudo-true parameter
\begin{align}
	 \label{eqn:DefinitionPseudoTrueParameter}
	\theta^\ast_T = \argmin_{\theta \in \Theta} Q_T^0(\theta), \qquad \text{ where } \qquad Q_T^0(\theta) = \mathbb{E} [ Q_T(\theta) ] 
\end{align}
For the classical case of a correctly specified model, the pseudo-true parameter coincides with the true regression parameter $\theta^\ast_T = \theta_0$ and is independent of $T$.
In the following, we restrict our attention to processes and models for the conditional quantile and ES which follow the following conditions.
\newcounter{ESConditionsCounter}
\begin{assumption}
	\label{ass:Assumption1ESReg}
	$ $
	\begin{enumerate}[label=(A\arabic*)]
		\item 
		\label{cond:ContinuousDistributionESReg}
		The distribution $F_t$ is absolutely continuous with density function $f_t$, which is bounded from above, i.e. there exists a constant $c > 0$ s.t. $\sup_{y \in \mathbb{R}} f_t(y) \le c$ and $\sup_{y \in \mathbb{R}} f_t'(y) \le c$.
		
		\item 
		\label{cond:CompactParameterSpaceESReg}
		The parameter space $\Theta \subseteq \mathbb{R}^{2k}$ is compact, convex and has non-empty interior.
		
		\item 
		\label{cond:UniqueMinimum}
		We assume that the pseudo-true parameter $\theta_T^\ast$ defined in (\ref{eqn:DefinitionPseudoTrueParameter}) is in the interior of $\Theta$ and is the unique minimizer of the objective function $Q_T^0(\theta)$ and that the sequence $\nabla_\theta \mathbb{E}_t [ \rho(Y_t,X_t,\theta^\ast_T)]$ is uncorrelated.
		
		
		
		\item 
		$V_t, W_t \in \mathcal{F}_{t-1}$ and the matrices $\mathbb{E}[V_t V_t^\top]$ and $\mathbb{E}[W_t W_t^\top]$ have full rank.
		
		\item 
		\label{cond:LambdaEigenvalues}
		The matrix $\Lambda_T$, defined in Theorem \ref{thm:AsymptoticNormalityMestimatorMisspecifiedModel} has strictly positive Eigenvalues for all $T$ sufficiently large enough.
		
%
%
		
		\item 
		\label{cond:StrongMixingESReg}
		The stochastic process $\{Y_t, V_t, W_t\}$ is strong mixing of size $-r/(r-2)$ for some $r>2$.
		
		%
		%
		%

		\item 
		\label{cond:ESModelBounded}
		For all $\theta \in \Theta$, it holds that $\left| \frac{1}{W_t^\top \gamma} \right| \le K < \infty$ for some constant $K > 0$.

		\item 
		\label{cond:MomentConditionsESReg}
		It holds that $\mathbb{E} \left[ ||V_t||^{r+1} \right] < \infty$, $\mathbb{E} \left[ ||W_t||^{r+1} \right] < \infty$, $\mathbb{E} \left[ ||V_t||^{r+1}  ||W_t||^{r}  \right] < \infty$ and $\mathbb{E} \left[ ||W_t||^{r+1}  |Y_t|^{r}  \right]< \infty$ for the $r > 2$ from condition \ref{cond:StrongMixingESReg}.

		\item 
		\label{cond:ExactHits}
		For any $T \in \mathbb{N}$, $\sup_{\theta \in \Theta} \mathds{1}_{\{ Y_t = V_t^\top \beta \}} \le K$ a.s. for some constant $K > 0$.

		%
	\end{enumerate}
\end{assumption}

The conditions in Assumption \ref{ass:Assumption1ESReg} mainly resemble the regularity conditions for asymptotic normality for correctly specified models from \cite{Patton2019} and we refer to \cite{Patton2019} for a discussion of these conditions.
The key condition which allows for misspecified models is the unique minimization condition of the pseudo-true parameter $\theta_T^\ast$ in condition \ref{cond:UniqueMinimum}.
The above assumptions contain the case of correctly specified models as then, the condition  \ref{cond:UniqueMinimum} is naturally fulfilled as the utilized loss function is a strictly consistent loss function for the VaR and the ES \citep{Fissler2016}.

We connect this weaker condition \ref{cond:UniqueMinimum} to classical misspecified regression models for the mean and for quantiles of \cite{White1980}, \cite{Gourieroux1984}, \cite{KimWhite2003}, \cite{Komunjer2005} and \cite{Angrist2006}.
For correctly specified models, we usually impose the strong condition that for all $t = 1,\dots, T$,
\begin{align}
	\mathbb{E}_t \big[ \psi(Y_t,X_t,\theta) \big] = 0 \quad \text{ a.s. } \qquad \Longleftrightarrow \qquad \theta = \theta_T^\ast,
\end{align}
where $\psi(Y_t,X_t,\theta)$ is almost surely the derivative of $\rho(Y_t,X_t,\theta)$ and corresponds to the identification functions of the model \citep{Gneiting2011}.
The weaker condition \ref{cond:UniqueMinimum} is essentially equivalent to the unconditional moment condition
\begin{align}
	\label{eqn:AverageIdentificationCondition}
	\mathbb{E} \left[ \frac{1}{T} \sum_{t=1}^T \psi(Y_t,X_t,\theta) \right] = 0 \qquad \Longleftrightarrow \qquad \theta = \theta_T^\ast.
\end{align}
Thus, the condition (\ref{eqn:AverageIdentificationCondition}) can be interpreted as an \textit{average identification} condition, i.e. $V_t^\top \beta_T^\ast$  and  $W_t^\top \gamma_T^\ast$ are some best averaged linear approximations of the true unknown conditional quantile and ES models.

\begin{theorem}[Consistency Misspecified Model]
	\label{thm:ConsistencyMestimatorMisspecifiedModel}
	Given the conditions from Assumption \ref{ass:Assumption1ESReg}, it holds that
	$\hat \theta_T - \theta^\ast_T \stackrel{\mathbb{P}}{\to} 0$,
	as $T \to \infty$, where $\theta^\ast_T$ is the pseudo-true parameter as defined in (\ref{eqn:DefinitionPseudoTrueParameter}).
\end{theorem}
The proof of Theorem \ref{thm:ConsistencyMestimatorMisspecifiedModel} is given in Appendix \ref{sec:Proofs}.

\begin{theorem}[Asymptotic Normality Misspecified Model]
	\label{thm:AsymptoticNormalityMestimatorMisspecifiedModel}
	Given the conditions of Assumption \ref{ass:Assumption1ESReg},	it holds that
	\begin{align}
	\Sigma_T(\theta_T^\ast)^{-1/2} \Lambda_T(\theta_T^\ast) \, \sqrt{T} \big(\hat \theta_T - \theta^\ast_T \big) \stackrel{d}{\to} \mathcal{N} \left( 0, I_{2k} \right),
	\end{align}
	where
	\begin{align}
		\Lambda_T(\theta^\ast_T) = 
		\begin{pmatrix}
		\Lambda_{11,T}(\theta^\ast_T)  & \Lambda_{12,T}(\theta^\ast_T)  \\
		\Lambda_{21,T}(\theta^\ast_T)  & \Lambda_{22,T}(\theta^\ast_T) 
		\end{pmatrix}
		\qquad \text{ and } \qquad
		\Sigma_T(\theta^\ast_T) = 
		\begin{pmatrix}
			\Sigma_{11,T}(\theta^\ast_T)  & \Sigma_{12,T}(\theta^\ast_T)  \\
			\Sigma_{21,T}(\theta^\ast_T)  & \Sigma_{22,T}(\theta^\ast_T) 
		\end{pmatrix}
	\end{align}
	with
	\begin{align}
		\label{eqn:Lambda11}
		\Lambda_{11,T}(\theta^\ast_T)  &= - \frac{1}{T} \sum_{t=1}^T\mathbb{E} \left[ V_t V_{t}^\top f_t(V_{t}^\top \beta^\ast_T) \frac{1}{\tau W_{t}^\top \gamma^\ast_T}  \right], \\
		\Lambda_{12,T}(\theta^\ast_T)  &= \Lambda_{21,T}^\top(\theta^\ast_T)  = \frac{1}{T} \sum_{t=1}^T\mathbb{E} \left[ V_t W_t^\top  \frac{1}{(W_t^\top \gamma^\ast_T)^2}  \frac{F_t(V_t^\top \beta^\ast_T) - \tau}{\tau} \right], \\
		\Lambda_{22,T}(\theta^\ast_T)  &= \frac{1}{T} \sum_{t=1}^T \mathbb{E} \left[ W_t W_t^\top \frac{1}{(W_t^\top \gamma^\ast_T)^2}  \right] \\
		& - \frac{2}{T} \sum_{t=1}^T \mathbb{E} \left[ W_t W_t^\top \frac{1}{(W_t^\top \gamma^\ast_T)^3} \left( W_t^\top \gamma^\ast_T - \frac{1}{\tau} \mathbb{E} _t \left[  Y_t \mathds{1}_{\{Y_t \le V_t^\top \beta^\ast_T \}} \right] +  V_t^\top \beta^\ast_T \frac{ F_t(V_t^\top \beta^\ast_T) - \tau}{\tau} \right)  \right],
	\end{align}
	and
	\begin{align}
		\Sigma_{11,T}(\theta^\ast_T)  &= \frac{1}{T}  \sum_{t=1}^T \mathbb{E} \left[ V_t V_t^\top \frac{1}{(W_t^\top \gamma^\ast_T)^2} \left( \frac{1-\tau}{\tau} + \frac{(1-2\tau) (F_t(V_t^\top \beta_T^\ast) - \tau)}{\tau^2}  \right) \right],  \\
		\Sigma_{12,T}(\theta^\ast_T)  &= \frac{1}{T}  \sum_{t=1}^T \mathbb{E} \left[ V_t W_t^\top \frac{1}{-(W_t^\top \gamma^\ast_T)^3}  \left\{ \frac{1-\tau}{\tau} \left( V_t^\top \beta_T^\ast  - W_t^\top \gamma_T^\ast \right) \right.  \right. \\
		& \qquad\qquad\qquad+ \frac{1-\tau}{\tau} \left( V_t^\top \beta_T^\ast \frac{F_t(V_t^\top \beta_T^\ast) - \tau}{\tau} + W_t^\top \gamma_T^\ast - \frac{1}{\tau} \mathbb{E}_t \left[ Y_t \mathds{1}_ {\{ Y_t \le V_t^\top \beta_T^\ast \}} \right] \right) \\
		& \qquad\qquad\qquad - \left. \left. \frac{F_t(V_t^\top \beta_T^\ast) - \tau}{\tau} \left( V_t^\top \beta_T^\ast  - W_t^\top \gamma_T^\ast \right) \right\} \right], \\ 
		\Sigma_{22,T}(\theta^\ast_T)  &= \frac{1}{T}  \sum_{t=1}^T \mathbb{E} \left[ W_t W_t^\top \frac{1}{(W_t^\top \gamma^\ast_T)^4} \left\{  \frac{1}{\tau} \operatorname{Var}_t(V_t^\top \beta_T^\ast-Y_t| Y_t \le V_t^\top \beta_T^\ast)  + \frac{1- \tau}{\tau} \big(V_t^\top \beta_T^\ast  - W_t^\top \gamma_T^\ast  \big)^2 \right. \right. \\
		& \qquad\qquad\qquad \left. \left.  + 2  \big(V_t^\top \beta_T^\ast  - W_t^\top \gamma_T^\ast  \big) V_t^\top \beta_T^\ast \frac{\tau - F_t(V_t^\top \beta_T^\ast) }{\tau} \right\} \right].
		\label{eqn:Sigma22}
	\end{align}

\end{theorem}
The proof of Theorem \ref{thm:AsymptoticNormalityMestimatorMisspecifiedModel} is given in Appendix \ref{sec:Proofs}.
The asymptotic theory derived here embeds the asymptotic theory of \cite{Patton2019} and \cite{DimiBayer2019} in the simplified case of correctly specified models.
Correct specification implies that $F_t(V_t^\top \beta^\ast_T) = \tau$ and $W_t^\top \gamma^\ast_T = \frac{1}{\tau} \mathbb{E} _t \big[ Y_t \mathds{1}_{\{Y_t \le V_t^\top \beta^\ast_T \}} \big]$ almost surely for all $t = 1,\dots,T$.
Imposing these two conditions simplifies the asymptotic covariance matrix of Theorem \ref{thm:AsymptoticNormalityMestimatorMisspecifiedModel} to the asymptotic covariances from \cite{Patton2019} and \cite{DimiBayer2019}.
In general, allowing for model misspecification in regression models comes at the cost of an inflated and more complicated asymptotic covariance matrix, see e.g. \cite{White1980}, \cite{White1994}, \cite{KimWhite2003}, \cite{Komunjer2005} and \cite{Angrist2006} for examples of semiparametric models for the mean and quantiles.

Given consistency and asymptotic normality, we can derive the asymptotic distribution of the test statistics of our new regression-based ESR backtests.
Henceforth, we use the short notation $\Omega_{T} = \Lambda_T(\theta_T^\ast)^{-1} \Sigma_T(\theta_T^\ast) \Lambda_T(\theta_T^\ast)^{-1}$ for the asymptotic covariance.
As the Auxiliary ESR backtest is not subject to model misspecification,  under the null hypothesis it holds that $\gamma_T^\ast = (0,1)$ for all $T \in \mathbb{N}$.
However, this does not necessarily hold for the Strict ESR and the Intercept ESR backtests and we consequently define the following modified test statistics for these backtests,
\begin{align}
	\widetilde T_{\text{S-ESR}} &= T \left( \hat \gamma_T - \gamma^\ast_T \right) \, \widehat \Omega_{T,\gamma}^{-1}  \,  \left( \hat \gamma_T -  \gamma^\ast_T \right)^\top,	 \\
	\widetilde T_{\text{I-ESR}} &= T \left( \hat \gamma_{1,T} - \gamma^\ast_{1,T} \right) \, \widehat \Omega_{T,\gamma_1}^{-1}  \,  \left( \hat \gamma_{1,T} -  \gamma^\ast_{1,T} \right)^\top,	
\end{align}	
where $\widehat \Omega_{T,\gamma}$ and $\widehat \Omega_{T,\gamma_1}$ are the ES-specific parts of the estimators for the asymptotic covariance matrix and $\gamma^\ast_{1,T}$ refers to the intercept component of the pseudo-true ES specific parameter vector $\gamma^\ast_{T}$.
\begin{corollary}
	\label{cor:TestStatisticChiSq}
	Given the conditions of Assumption \ref{ass:Assumption1ESReg} and given that 
	$\widehat \Omega_{T} - \Omega_{T} \stackrel{\mathbb{P}}{\to} 0$, it holds that
	\begin{align}
		T_{\text{A-ESR}} \stackrel{d}{\to} \chi^2_2, \qquad 
		\widetilde T_{\text{S-ESR}} \stackrel{d}{\to} \chi^2_2, \qquad \text{ and } \qquad
		\widetilde T_{\text{I-ESR}} \stackrel{d}{\to} \chi^2_1.
	\end{align}
\end{corollary}
The proof of Corollary \ref{cor:TestStatisticChiSq} is given in Appendix \ref{sec:Proofs}.
For the Strict ESR test (and the intercept version), we do not know the exact form of the peuso-true parameter $\gamma_T^\ast$ in practice.
In the following, we argue that in realistic financial settings, $\gamma_T^\ast \approx (0,1)$ and thus, $\widetilde T_{\text{S-ESR}} \approx T_{\text{S-ESR}}$ holds approximately.
First, the majority of literature in financial econometrics finds that pure scale processes (e.g. GARCH and stochastic volatility models) approximate the true underlying daily financial data well enough.
Thus, $\hat v_{t} \approx c \hat e_t$ for some $c > 0$ and we find that under the null hypothesis, the regression model in (\ref{eqn:RegSpecStrictES}) is only subject to a \textit{slight} model misspecification.
Second, the misspecification is in the \textit{auxiliary} quantile equation, while we test the parameters of the correctly specified ES equation in (\ref{eqn:RegSpecStrictES}).
Thus, the model misspecification enters our test statistic only indirectly through the auxiliary effect of the joint parameter estimation.
Third, our simulation results in Section \ref{sec:monte_carlo} show that the Strict ESR backtest based on $T_{\text{S-ESR}}$ exhibits correct size properties and performs almost indistinguishably to the Auxiliary ESR backtest, also in the simulation setups where the underlying data does not follow a pure scale processes.
This shows that the approximation error is negligible in realistic financial settings and that the Strict and Intercept ESR backtests can be applied in practice.

%
%
%
%
%
%
%
%
%
%
%

\subsection{Implementation of the Tests}
\label{sec:TestImplementation}

The M-estimation of the parameters $\hat \theta_T$ is carried out by using the R package \texttt{esreg} \citep{BayerDimi2019esreg}.
The main difficulty in the implementation of the backtests is estimation of the asymptotic covariance matrix $\Omega_{T} = \Lambda_T(\theta_T^\ast)^{-1} \Sigma_T(\theta_T^\ast) \Lambda_T(\theta_T^\ast)^{-1}$.
Generally, this is implemented by using the sample counterparts of the expectation of the components given in (\ref{eqn:Lambda11}) - (\ref{eqn:Sigma22}) in Theorem \ref{thm:AsymptoticNormalityMestimatorMisspecifiedModel}, wich are however subject to the following four nuisance quantities:
\begin{enumerate}[label = (\alph*)]
	\item
	the conditional density function, evaluated at the conditional quantile, $\hat f_t(V_t^\top \hat \beta_T)$,
	
	\item	
	the conditional, truncated variance, $\widehat{\operatorname{Var}}_t(V_t^\top \hat \beta_T-Y_t| Y_t \le V_t^\top \hat \beta_T)$,
	
	\item
	the conditional distribution function, $\hat F_t(V_t^\top \hat \beta_T)$, and
	
	\item
	the conditional, truncated expectation  $\frac{1}{\tau} \mathbb{E} _t \left[  Y_t \mathds{1}_{\{Y_t \le V_t^\top \hat \beta_T \}} \right]$.
	
\end{enumerate}

We implement a novel and misspecification robust covariance estimator by estimating the four nuisance quantities above in the following way.
The terms (a) and (b) are subject to the asymptotic covariance of correctly specified models for the quantile and the ES of  \cite{DimiBayer2019}, \cite{Patton2019} and \cite{Barendse2017}.
Thus, we follow the approach of \cite{DimiBayer2019} and apply the \textit{nid} estimator of \cite{Hendricks1992} for (a), the conditional density and the flexible \textit{scl-sp} estimator of \cite{DimiBayer2019} for (b), the conditional truncated variance.

%
%
%

In order to estimate (c), the conditional distribution function $\hat F_t(V_t^\top \hat \beta_T)$, we follow the general approach of the \textit{scl-sp} estimator of \cite{DimiBayer2019}, i.e. we assume that $F_t$ follows a conditional location-scale model with innovations $\varepsilon_t$ with a flexible zero mean and unit variance distribution.
We standardize $Y_t$ by the estimates of the conditional mean and variance, estimated by pseudo-maximum likelihood and apply a kernel density estimator in order to obtain the distribution function of $\varepsilon_t$.
Hence, we can recover the distribution of $Y_t$ given $\mathcal{F}_{t-1}$.
Notice that for the minor degree of misspecification we are subject to in our backtesting approach, it approximately hold that $\hat F_t(V_t^\top \hat \beta_T) \approx \tau$ for all $t$.
We find that this semiparametric estimation approach, which is subject to the location-scale assumption, performs better than pure nonparametric alternatives as we are estimating the conditional distribution evaluated at rather extreme quantiles such as at $\tau = 2.5\%$.

The last nuisance quantity, $\frac{1}{\tau} \mathbb{E} _t \left[  Y_t \mathds{1}_{\{Y_t \le V_t^\top \hat \beta_T \}} \right]$, is the mean, given the observations are smaller than the possibly misspecified linear quantile model. 
This quantity is closely related the the conditional ES, which is  assumed to be a linear function in our approach.
As for realistic financial data, we only face a minor degree of misspecification in the quantile model, this nuisance quantity is assumed to still be approximately linear, and thus, we obtain that $\frac{1}{\tau} \mathbb{E} _t \left[  Y_t \mathds{1}_{\{Y_t \le V_t^\top \hat \beta_T \}} \right] = W_t^\top \hat \gamma_t$ for all $t$.
Nonparametric estimation of this nuisance quantity again introduces too much estimation noise.

We further implement our backtests based on a covariance estimator from \cite{DimiBayer2019} and \cite{Patton2019}, which does not account for possible model misspecification.
This estimator is based on the simplified covariance structure given in \cite{DimiBayer2019} and \cite{Patton2019}, where the correct model specification assumption implies that $F_t(V_t^\top \beta_T^\ast) = \tau $, and $\frac{1}{\tau} \mathbb{E} _t \left[  Y_t \mathds{1}_{\{Y_t \le V_t^\top \beta_T^\ast \}} \right] = W_t^\top \gamma_T^\ast$ almost surely.
Thus, we only estimate the nuisance quantities (a) and (b) in this approach.

\section{Monte-Carlo Simulations}
\label{sec:monte_carlo}

In this section, we evaluate the empirical performance of our proposed ESR backtests and compare them to the tests of \citet{McNeil2000} and \citet{Nolde2017}.
For that, we assess the empirical size and power of the tests, which are defined as the rejection frequency of the tests under the null and alternative hypothesis respectively.
This comparison is conducted using two different approaches.
The first, presented in \Cref{sec::TraditionalSizePower}, follows the typical strategy in the related literature of first assessing the size of the backtests with several realistic data generating processes (DGP), followed by an evaluation of the power by backtesting forecasts stemming from an overly simplified model, in this case the Historical Simulation (HS) model.
In the second setup, presented in \Cref{sec::ContinuousMisspecification}, we continuously misspecify certain parameters of the true model and thereby obtain alternative models with a continuously increasing degree of misspecification.
This approach of evaluating backtests has two advantages.
First, we obtain power curves which can be used to draw conclusions how an increasing model misspecification influences the test decisions.
Second, misspecifying the different model parameters in isolation allows us to misspecify certain model characteristics while leaving the remaining model unchanged.

\subsection{Traditional Size and Power Comparisons}
\label{sec::TraditionalSizePower}

In order to compare the proposed backtests from the previous sections, we simulate data from several DGPs.
Besides pure scale (volatility) model specifications, under which the Strict and Intercept ESR backtests are correctly specified, we also consider more general Student's-$t$  GAS models \citep{Creal2013} with time-varying higher moments and AR-GARCH specifications where our ESR backtests are subject to model misspecification under the null hypothesis.


\begin{enumerate}[label=DGP-(\arabic*), labelindent=0em, labelsep=0.1cm, leftmargin=1cm]
	\item[\textbf{EGARCH:}]
	\label{model:EGARCH}
	The first DGP is an EGARCH(1,1) model \citep{Nelson1991} with $t$-distributed innovations, where the parameter values are calibrated to  daily returns of the S\&P\,500 index,
	\begin{align}
		\begin{aligned}
		\label{eq:sim_1_dgp}
		Y_t  & = \sigma_t z_t, \quad \text{ where } \quad	z_t \sim t_{7.39}, \quad \text{ and }  \\
		\log(\sigma_t^2) & = -0.0012 -0.161 z_{t-1} + 0.136 \left(|z_{t-1}| - E [|z_{t-1}|] \right) + 0.978 \log (\sigma_{t-1}^2 ),
		\end{aligned}
	\end{align}
	This model represents a highly flexible GARCH specification and due to its calibrated parameter values, this DGP accurately replicates the distributional properties of daily financial returns.
	As we assume a zero mean for this model, the true VaR and ES forecasts are perfectly colinear and consequently, the regression equations for the Strict and the Intercept ESR backtests are correctly specified under the null hypothesis.
	
	\item[\textbf{AR-GARCH:}]
	\label{model:ARGARCH}
	The next specification is an AR(1)-GARCH(1,1) model with Gaussian innovations,
	\begin{align}
		\begin{aligned}
		Y_t & = \phi Y_{t-1} + \sigma_t z_t, \quad \text{ where } \quad	z_t \sim \mathcal{N}(0,1), \quad \text{ and }  \\
		\sigma_t^2 & = 0.01 + 0.1 Y_{t-1}^2 + 0.85\sigma_{t-1}^2, 
		\end{aligned}
	\end{align}
	where we consider the three specifications $\phi \in \{ 0, 0.1, 0.5\}$ for the AR parameter.
	This DGP introduces model misspecification for the Strict and Intercept ESR backtests through the non-zero conditional mean specification, while leaving the realistic volatility structure of the financial returns unchanged.
	For this DGP, the ratio between true VaR and ES is given by
	\begin{align}
		\label{eqn:ModelMisspARGARCH}
		\frac{\hat v_t}{\hat e_t} = \frac{\mu_t +  \sigma_t z_\tau}{\mu_t +  \sigma_t \xi_\tau},
	\end{align}
	where $\mu_t$ is the conditional mean of $Y_t$ given $\mathcal{F}_{t-1}$.
	If $\mu_t$ equals zero, the ratio is constant and thus, the regression equations in (\ref{eqn:RegSpecStrictES}) are correctly specified under the null.
	By increasing the time-dependence of the conditional mean model through the AR parameter, we can monotonically strengthen the model misspecification in this DGP.
	
	\item[\textbf{GAS-STD:}]
	\label{model:GAS3Factor}
	We use a 3-factor Student's-$t$ GAS model with time-varying location $\mu_t$, scale $\sigma_t$, and degrees of freedom $\xi_t$ with parameters calibrated to daily returns of the S\&P\,500 index.
	This model is estimated and simulated by using the R package GAS \citep{Ardia2019GAS} and is based on the following model specification
	\begin{align}
		\label{eqn:GASModel3FactorT}
		Y_t \big| \big( Y_1, \dots,Y_{t-1} \big) &\sim t ( \mu_t, \sigma_t, \xi_t ), 
	\end{align}
	where the vector $(\mu_t, \sigma_t, \xi_t)$ follows an autoregressive specification, driven by the lagged score of the log-likelihood of the distributional specification in (\ref{eqn:GASModel3FactorT}).
	\cite{Creal2013} and \cite{Harvey2013} introduce the general GAS specification, which nests many well known models, including ARMA, GARCH \citep{Bollerslev1986} and ACD \citep{Engle1998} models. 
	\cite{Koopman2016} provides an overview of GAS and related models.
	We refer to Appendix A of \cite{Ardia2019GAS} for the exact parametric specification of this Student's-$t$ GAS model.

	\item[\textbf{GAS-SSTD:}]
	\label{model:GAS4Factor}
	We generalize the previous GAS model to a 4-factor asymmetric Student's-$t$ GAS model with time-varying location $\mu_t$, scale $\sigma_t$, skewness $\lambda_t$, and degrees of freedom $\xi_t$,
	\begin{align}
		\label{eqn:GASModel4FactorT}
		Y_t \big| \big( Y_1, \dots,Y_{t-1} \big) &\sim t ( \mu_t, \sigma_t, \lambda_t, \xi_t ).
	\end{align}
	Compared to the previous 3-factor GAS specification, this model further allows for asymmetries in the conditional return distribution through allowing for an additional time-varying skewness parameter with an autoregressive GAS-specification.
\end{enumerate}

For the two location-scale DGPs, we obtain VaR and ES forecasts at level $\tau$ by
\begin{align}
	\label{eq:sim_1_true}
    \hat{v}_t = \hat \mu_t + \hat \sigma_t q_z(\tau) \qquad \text{and} \qquad \hat{e}_t = \hat \mu_t + \hat \sigma_t \xi_z(\tau),
\end{align}
where $\hat \mu_t$ and $\hat \sigma_t$ are the respective location and volatility forecasts generated by the location and scale models and $q_z(\tau)$ and $\xi_z(\tau)$ are the $\tau$-quantile, respectively the $\tau$-ES of the innovations $z_t$.
For the $t$-distributions of the two GAS models, we obtain the ES forecasts through numerical integration.
For the following size and power analysis of the backtests, we simulate data from the DGPs given above with varying sample sizes of 250, 500, 1000, 2500, and 5000 observations and 250 additional pre-sample values required for the power analysis.
We run 10,000 Monte Carlo replications for each of the DGPs.
As stipulated by the Basel Accords, we fix the probability level to $\tau=2.5\%$ for the VaR and ES forecasts for each of the DGPs.
In this part of the study, we focus on two-sided hypotheses and defer the one-sided case to \Cref{sec:one_sided_alternative}.
We compare our three ESR backtests to two specifications of the conditional calibration (CC) backtest of \cite{Nolde2017} and to two specifications of the exceedance residual (ER) backtests of \cite{McNeil2000}, which are further described in Appendix \ref{sec:er_test} and Appendix \ref{sec:cc_test}.

\Cref{tab:mc1_size1} presents the empirical sizes of the considered backtests for the different DGPs introduced above and for the different sample sizes and a nominal test size of 5\%.
\Cref{tab:mc1_size1Percent} and \Cref{tab:mc1_size10Percent} in Appendix \ref{sec:AdditionalMaterial} show equivalent results for nominal significance levels of 1\% and  and 10\%.
We find that in large samples, all backtests display rejection rates close to the respective nominal size for all considered DGPs.
However, in small samples the ESR tests based on the misspecification covariance estimator exhibit much better sizes compared to the equivalent ESR tests which do not account for the potential misspecification.
As this holds for both, DGPs which do and do not generate misspecification under the null, this indicates that the misspecification covariance estimator better approximates the finite sample distribution and should consequently be applied in empirical applications.

We further find that the Strict ESR test and the Auxiliary ESR test perform very similar throughout all considered DGPs.
This implies that the indirect misspecification the Strict ESR test introduces is negligible for realistic financial data.
Even for the AR-GARCH model with increasing AR paramter $\phi$, the size properties of the Strict and the Intercept ESR tests are not adversely affected by the increasing degree of misspecification.
From the four competitor backtests, the general CC and the ER and its standardized version exhibit satisfactory sizes whereas the Simple CC test is severely oversized, especially in small samples.

\begin{table}
	\footnotesize
	\centering
	\caption{Empirical sizes for the first simulation study.}
	\label{tab:mc1_size1}
	\begin{tabularx}{\linewidth}{XX *{12}{r}}
		\toprule
		\theadl{DGP} & \theadl{Sample \\ Size} & \thead{Str. \\ ESR} & \thead{Aux. \\ ESR} & \thead{Int. \\ ESR} & \thead{Str. \\ ESR} & \thead{Aux. \\ ESR} & \thead{Int. \\ ESR}  & \thead{Gen. \\ CC} & \thead{Sim. \\ CC} & \thead{Std. \\ ER} & \thead{ER} \\
		\cmidrule(lr){3-5} \cmidrule(lr){6-8} \cmidrule(lr){9-12}
		& & \multicolumn{3}{c}{Misspec Covariance} & \multicolumn{3}{c}{Classical Covariance} \\	
		\midrule
            &   250 & 0.09 & 0.09 & 0.14 & 0.24 & 0.25 & 0.16 & 0.08 & 0.29 & 0.07 & 0.09 \\
            &   500 & 0.06 & 0.07 & 0.10 & 0.15 & 0.15 & 0.11 & 0.10 & 0.20 & 0.04 & 0.07 \\
 EGARCH-STD &  1000 & 0.05 & 0.05 & 0.08 & 0.11 & 0.11 & 0.08 & 0.09 & 0.14 & 0.05 & 0.07 \\
            &  2500 & 0.04 & 0.04 & 0.06 & 0.06 & 0.06 & 0.06 & 0.07 & 0.09 & 0.05 & 0.06 \\
            &  5000 & 0.04 & 0.04 & 0.07 & 0.06 & 0.06 & 0.07 & 0.06 & 0.08 & 0.05 & 0.06 \\
		\midrule
         &   250 & 0.10 & 0.10 & 0.14 & 0.26 & 0.26 & 0.15 & 0.07 & 0.28 & 0.07 & 0.08 \\
         &   500 & 0.07 & 0.08 & 0.10 & 0.16 & 0.16 & 0.11 & 0.10 & 0.20 & 0.06 & 0.06 \\
 GAS-STD &  1000 & 0.06 & 0.06 & 0.07 & 0.11 & 0.11 & 0.07 & 0.09 & 0.14 & 0.06 & 0.07 \\
         &  2500 & 0.05 & 0.05 & 0.06 & 0.08 & 0.08 & 0.06 & 0.07 & 0.10 & 0.06 & 0.06 \\
         &  5000 & 0.04 & 0.05 & 0.08 & 0.06 & 0.06 & 0.08 & 0.06 & 0.08 & 0.06 & 0.06 \\
		\midrule
          &   250 & 0.09 & 0.09 & 0.13 & 0.25 & 0.25 & 0.15 & 0.07 & 0.26 & 0.08 & 0.07 \\
          &   500 & 0.06 & 0.06 & 0.10 & 0.15 & 0.15 & 0.10 & 0.09 & 0.18 & 0.06 & 0.05 \\
 GAS-SSTD &  1000 & 0.05 & 0.05 & 0.07 & 0.10 & 0.10 & 0.07 & 0.08 & 0.13 & 0.07 & 0.06 \\
          &  2500 & 0.04 & 0.04 & 0.06 & 0.06 & 0.06 & 0.06 & 0.07 & 0.09 & 0.06 & 0.05 \\
          &  5000 & 0.04 & 0.04 & 0.06 & 0.05 & 0.05 & 0.06 & 0.07 & 0.07 & 0.06 & 0.05 \\
		\midrule
                      &   250 & 0.05 & 0.04 & 0.11 & 0.18 & 0.18 & 0.13 & 0.06 & 0.22 & 0.06 & 0.07 \\
                      &   500 & 0.04 & 0.04 & 0.09 & 0.12 & 0.12 & 0.09 & 0.07 & 0.14 & 0.04 & 0.04 \\
 AR-GARCH, $\phi=0.0$ &  1000 & 0.03 & 0.04 & 0.07 & 0.09 & 0.09 & 0.07 & 0.07 & 0.10 & 0.04 & 0.04 \\
                      &  2500 & 0.03 & 0.03 & 0.06 & 0.06 & 0.06 & 0.06 & 0.06 & 0.08 & 0.05 & 0.05 \\
                      &  5000 & 0.04 & 0.04 & 0.05 & 0.06 & 0.06 & 0.05 & 0.05 & 0.06 & 0.05 & 0.05 \\
		\midrule
                      &   250 & 0.05 & 0.05 & 0.11 & 0.18 & 0.18 & 0.13 & 0.06 & 0.22 & 0.06 & 0.07 \\
                      &   500 & 0.04 & 0.04 & 0.09 & 0.12 & 0.12 & 0.09 & 0.07 & 0.14 & 0.04 & 0.04 \\
 AR-GARCH, $\phi=0.1$ &  1000 & 0.04 & 0.04 & 0.07 & 0.09 & 0.09 & 0.07 & 0.07 & 0.10 & 0.04 & 0.04 \\
                      &  2500 & 0.03 & 0.03 & 0.06 & 0.07 & 0.07 & 0.06 & 0.06 & 0.08 & 0.05 & 0.05 \\
                      &  5000 & 0.04 & 0.04 & 0.05 & 0.06 & 0.06 & 0.05 & 0.05 & 0.06 & 0.05 & 0.05 \\
		\midrule
                      &   250 & 0.04 & 0.04 & 0.11 & 0.17 & 0.17 & 0.13 & 0.06 & 0.22 & 0.06 & 0.07 \\
                      &   500 & 0.04 & 0.04 & 0.09 & 0.12 & 0.12 & 0.09 & 0.07 & 0.14 & 0.04 & 0.04 \\
 AR-GARCH, $\phi=0.5$ &  1000 & 0.04 & 0.04 & 0.07 & 0.09 & 0.09 & 0.07 & 0.07 & 0.10 & 0.04 & 0.04 \\
                      &  2500 & 0.04 & 0.04 & 0.06 & 0.07 & 0.07 & 0.06 & 0.06 & 0.08 & 0.05 & 0.05 \\
                      &  5000 & 0.04 & 0.04 & 0.05 & 0.06 & 0.06 & 0.05 & 0.05 & 0.06 & 0.05 & 0.05 \\
		\bottomrule
		\addlinespace
        \multicolumn{12}{p{.97\linewidth}}{\textit{Notes:} The table reports the empirical sizes of the backtests for the different DGPs decribed in Section \ref{sec::TraditionalSizePower} and for a nominal test size of $5\%$.
		The number of Monte-Carlo repetitions is 10,000 and the probability level for the risk measures is $\tau=2.5\%$. 
		ESR refers to the three backtests introduced in this paper and we consider versions with covariance estimation with and without model misspecification.
		CC refers to the conditional calibration tests of \citet{Nolde2017}, and ER to the exceedance residuals tests of \citet{McNeil2000}.}
	\end{tabularx}
\end{table}

For a comparison of the power of the backtests, we evaluate their ability to reject the null hypothesis for risk models producing incorrect ES forecasts.
We utilize the Historical Simulation (HS) approach which forecasts the VaR and ES by using their empirical counterparts from previous trading days,
\begin{align}
	\hat{v}_t = \widehat{Q}_{\tau} \left(Y_{t-1}, \, Y_{t-2}, \cdots, Y_{t-w}\right)  \quad \text{and} \quad                                                                                           
	\hat{e}_t = \frac{1}{\sum_{i=1}^{w} \mathds{1}_{\left\{Y_{t-i} \leq \hat{v}_{t-i}\right\}}} \sum_{i=1}^{w} Y_{t-i} \cdot \mathds{1}_{\left\{Y_{t-i} \leq \hat{v}_{t-i} \right\}},
\end{align}
where $\widehat{Q}_{\tau}$ is the empirical $\tau$-quantile and $w$ is the length of a rolling window, that we set to 250, i.e. one year of data.
Since the standardized ER and the general CC backtests require forecasts of the volatility, we estimate this quantity with the sample standard deviation of the returns over the same rolling window.
For a meaningful and fair comparison of the power of the backtests to reject the null hypothesis, we compare the \textit{size-adjusted power}%
\footnote{%
    A comparison of the \textit{raw power}, i.e. the raw rejection rate of the null hypotheses, could be misleading due to the differences in the empirical sizes of the backtests.
    In particular, an oversized test would exhibit unrealistically large rejection rates.
}
of the backtests \citep{Lloyd2005}.
For this, the original critical values of the tests are modified such that the rejection frequencies of the true model equal the nominal test sizes.
The size-adjusted power is then given by the rejection frequencies of the alternative models using these modified critical values.

\begin{figure}
	\begin{subfigure}{.5\linewidth}
		\caption{EGARCH: Size-adjusted Power}
		\includegraphics{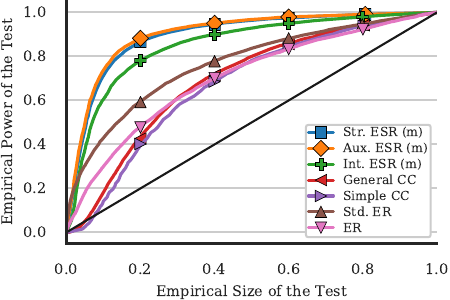}
		\label{fig:EGARCH_roc_plots}
	\end{subfigure}
	\begin{subfigure}{.5\linewidth}
		\caption{EGARCH: Partial Area Under the Curve}		
		\includegraphics{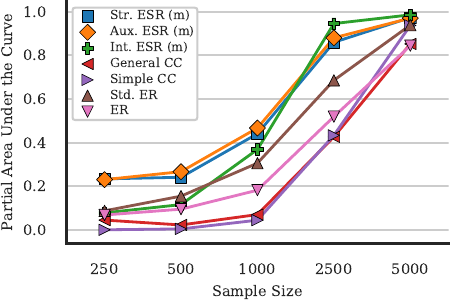}
		\label{fig:EGARCH_pauc_plots}
	\end{subfigure}
	\begin{subfigure}{.5\linewidth}
		\caption{GAS-STD: Size-adjusted Power}		
		\includegraphics{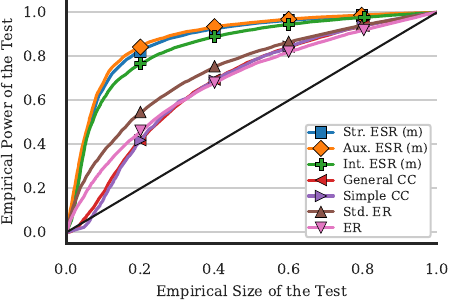}
		\label{fig:GASSTD_roc_plots}
	\end{subfigure}
	\begin{subfigure}{.5\linewidth}
		\caption{GAS-STD: Partial Area Under the Curve}		
		\includegraphics{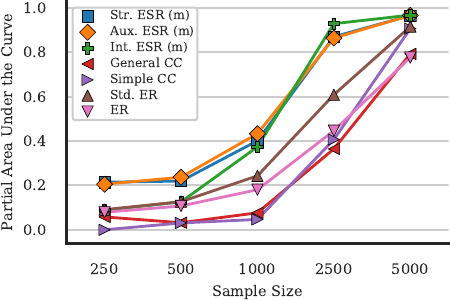}
		\label{fig:GASSTD_pauc_plots}
	\end{subfigure}
	\begin{subfigure}{.5\linewidth}
		\caption{GAS-SSTD: Size-adjusted Power}		
		\includegraphics{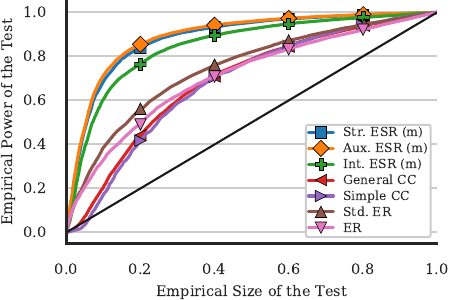}
		\label{fig:GASSSTD_roc_plots}
	\end{subfigure}
	\begin{subfigure}{.5\linewidth}
		\caption{GAS-SSTD: Partial Area Under the Curve}
		\includegraphics{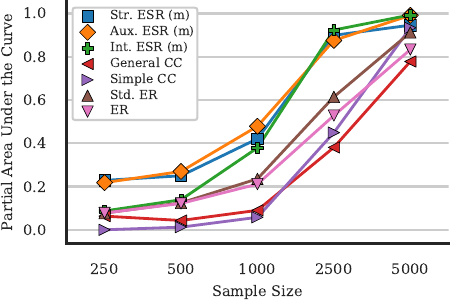}
		\label{fig:GASSSTD_pauc_plots}
	\end{subfigure}
	\caption[Power Plots EGARCH and GAS Models]{Size-adjusted power and Partial Area Under the Curve plots against Historical Simulation for a sample size of 1000 days.
	The number of Monte-Carlo repetitions is 10,000 and the probability level for the risk measures is $\tau=2.5\%$. ESR refers to the backtests introduced in this paper with (m) indicating the version which account for the additional covariance terms induced by the misspecified model. CC refers to the conditional calibration tests of \citet{Nolde2017}, and ER to the exceedance residuals tests of \citet{McNeil2000}.
}
	\label{fig:MC1_power_plots}
\end{figure}

\begin{figure}
	\begin{subfigure}{.5\linewidth}
		\caption{AR-GARCH $\phi = 0$: Size-adjusted Power}		
		\includegraphics{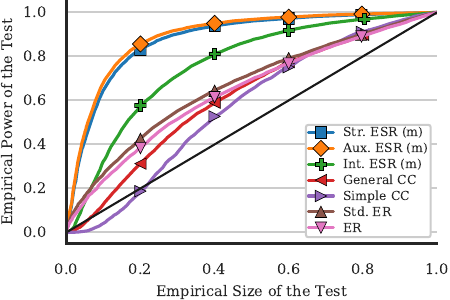}
		\label{fig:ARGARCH00_roc_plot}
	\end{subfigure}
	\begin{subfigure}{.5\linewidth}
		\caption{AR-GARCH $\phi = 0$: Partial Area Under the Curve}		
		\includegraphics{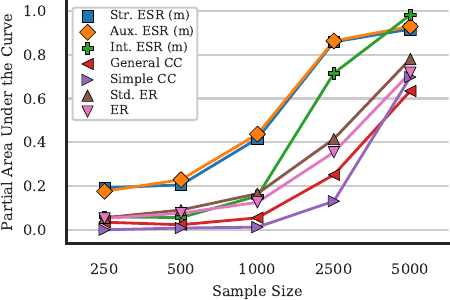}
		\label{fig:ARGARCH00_pauc_plot}
	\end{subfigure}
	\begin{subfigure}{.5\linewidth}
		\caption{AR-GARCH $\phi = 0.1$: Size-adjusted Power}		
		\includegraphics{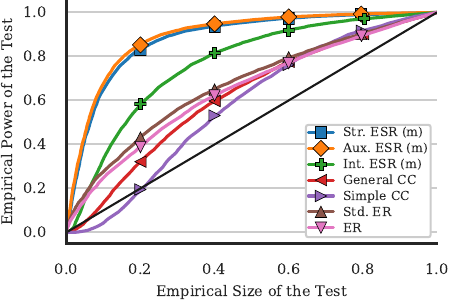}
		\label{fig:ARGARCH01_roc_plot}
	\end{subfigure}
	\begin{subfigure}{.5\linewidth}
		\caption{AR-GARCH $\phi = 0.1$: Partial Area Under the Curve}		
		\includegraphics{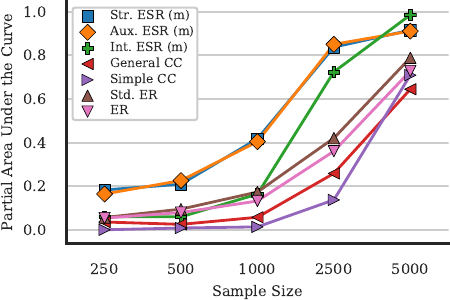}
		\label{fig:ARGARCH01_pauc_plot}
	\end{subfigure}
	%
	%
	\begin{subfigure}{.5\linewidth}
		\caption{AR-GARCH $\phi = 0.5$: Size-adjusted Power}
		\includegraphics{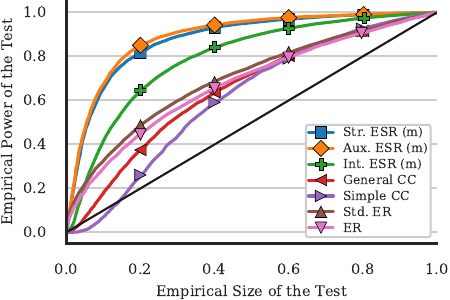}
		\label{fig:ARGARCH05_roc_plot}
	\end{subfigure}
	\begin{subfigure}{.5\linewidth}
		\caption{AR-GARCH $\phi = 0.5$: Partial Area Under the Curve}
		\includegraphics{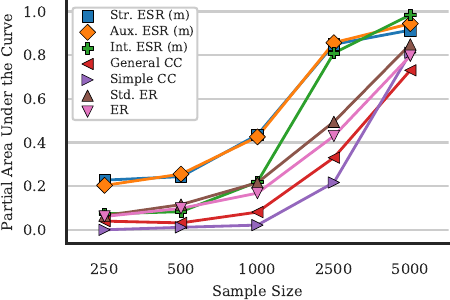}
		\label{fig:ARGARCH05_pauc_plot}
	\end{subfigure}
	\caption[Power Plots AR-GARCH models]{Size-adjusted power and Partial Area Under the Curve plots against Historical Simulation for a sample size of 1000 days.
	The number of Monte-Carlo repetitions is 10,000 and the probability level for the risk measures is $\tau=2.5\%$. ESR refers to the backtests introduced in this paper with (m) indicating the version which account for the additional covariance terms induced by the misspecified model. CC refers to the conditional calibration tests of \citet{Nolde2017}, and ER to the exceedance residuals tests of \citet{McNeil2000}.
	}
	\label{fig:MC1_power_plots_ARGARCH}
\end{figure}

The left panels in \Cref{fig:MC1_power_plots} and \Cref{fig:MC1_power_plots_ARGARCH} contain the size-adjusted power of the backtests for all empirical sizes in the unit interval for a sample size of 1000 and for the different DGPs.%
\footnote{
These plots are known as the receiver operating characteristic (ROC) curves and origin from the psychometrics literature \citep{Lloyd2005}.
They are an effective presentation method for general binary classification tasks such as hypothesis testing as they show the size-adjusted power simultaneously for all significance levels.
}
The black line depicts the case of equal empirical size and power, which can be seen as a lower bound for any reasonable test: whenever the power is below this line, randomly guessing the test decision is more accurate than performing the test.
For the three ESR backtests, we only report power for the tests relying on the misspecification robust covariance estimator as these versions of the tests exhibit superior size properties for all considered DGPs.
We observe that throughout all six considered DGPs, the three ESR backtests clearly dominate the four competitors in terms of power at almost all empirical sizes, including the most relevant region of test sizes between 1\% and 10\%.
Especially the Strict and the Auxiliary ESR tests exhibit a substantially larger power.

In order to present results for all considered sample sizes in condensed form for the relevant area of empirical sizes between 1\% and 10\%, we summarize the size-adjusted power by the partial area under the curve (PAUC), as proposed by \citet{Lloyd2005}.
For that, we numerically compute the area under each power curve for the empirical sizes between 1\% and 10\%, which can be interpreted as the test power averaged over the different test sizes.
In the right-hand panels of \Cref{fig:MC1_power_plots} and \Cref{fig:MC1_power_plots_ARGARCH}, we present the PAUC for all backtests, DGPs and sample sizes.
As expected, the average power increases with the sample size, so that using more information leads to more reliable decisions about the quality of a forecast.
We find that for all considered sample sizes, the ESR backtests dominate the other testing approaches.
This dominance is especially pronounced for the Strict and the Auxiliary ESR tests.
The almost identical performance of the Strict and the Auxiliary ESR tests throughout all simulation designs in \Cref{fig:MC1_power_plots} and \Cref{fig:MC1_power_plots_ARGARCH} emphasizes that the misspecification introduced by the Strict ESR test seems to be unproblematic for realistic financial data.


\subsection{Continuous Model Misspecification}
\label{sec::ContinuousMisspecification}

In the second simulation study, we use a GARCH(1,1) model with standardized Student-$t$ distributed innovations,
\begin{align}\label{eq:mc2_model}
\begin{split}
	Y_t        & =\sigma_t z_t,  \quad \text{ where } \quad		z_t    \sim t_{\nu}, \quad \text{ and }  \\    
	\sigma_t^2 & = \gamma_0 + \gamma_1 Y_{t-1}^2 + \gamma_2 \sigma_{t-1}^2,
\end{split}
\end{align}
with the parameter values $\gamma_0 = 0.01$, $\gamma_1 = 0.1$, $\gamma_2 = 0.85$, and $\nu=5$ for the true model.
For the analysis of the backtests, we simulate 10,000 times from this model with a fixed sample size of 2500 observations and consider the probability level $\tau=2.5\%$ for the VaR and the ES.
\begin{table}
	\footnotesize\centering
	\caption{Empirical sizes for the second simulation study.}
	\label{tab:mc2_size}
	\begin{tabularx}{\linewidth}{X *{11}{r}}
		\toprule
		\theadl{DGP} & \thead{Str. \\ ESR} & \thead{Aux. \\ ESR} & \thead{Int. \\ ESR} & \thead{Str. \\ ESR} & \thead{Aux. \\ ESR} & \thead{Int. \\ ESR}  & \thead{General \\ CC} & \thead{Simple \\ CC} & \thead{Std. \\ ER} & \thead{ER} \\
		\cmidrule(lr){2-4} \cmidrule(lr){5-7} \cmidrule(lr){8-11}
		& \multicolumn{3}{c}{Misspec Covariance} & \multicolumn{3}{c}{Classical Covariance} \\	
		\midrule
	 	Two-Sided & 0.07 & 0.07 & 0.06 & 0.05 & 0.05 & 0.06 & 0.07 & 0.09 & 0.05 & 0.05 \\
		One-Sided &   -- &   -- & 0.03 &   -- &   -- & 0.03 & 0.02 & 0.03 & 0.06 & 0.06 \\            
		\bottomrule
        \addlinespace
        \multicolumn{11}{p{.97\linewidth}}{\textit{Notes:} This table shows the empirical sizes of the backtests for the GARCH(1,1)-$t$ model given in (\ref{eq:mc2_model}), for a nominal test size of 5\% and for both, one-sided and two-sided hypotheses. 
        The number of Monte-Carlo repetitions is 10,000 and the probability level for the risk measures is $\tau=2.5\%$. ESR refers to the backtests introduced in this paper. CC refers to the conditional calibration tests of \citet{Nolde2017}, and ER to the exceedance residuals tests of \citet{McNeil2000}.
		Note that the Strict and Auxiliary ESR tests do not permit testing against a one-sided alternative and therefore, we only present sizes for the two-sided hypothesis.
		}
    \end{tabularx}
\end{table}
\Cref{tab:mc2_size} presents the empirical sizes of the backtests for a nominal size of 5\% for both, the two- and one-sided hypotheses.
As in the first simulation study, we find that most of the backtests are reasonably sized with rejection frequencies close to the nominal value.

For a detailed analysis of the power of the backtests, we continuously misspecify the true model according to the following five designs:

\begin{enumerate}[label=(\alph*), labelindent=0em, labelsep=0.1cm, leftmargin=*]
	\item 
	We misspecify how the conditional variance reacts to the squared returns by varying the ARCH parameter $\gamma_1$.	
	We choose $\tilde{\gamma}_1$ between 0.03 and 0.2 and let $\tilde{\gamma}_2 = 0.95 - \tilde{\gamma}_1$, such that the persistence of the GARCH process remains constant.
	When $\tilde{\gamma}_1 < {\gamma}_1$, there is too little variation in the ES forecasts due to the reduced response to shocks and the GARCH process approaches a constant volatility model.
	
	\item
	We alter the unconditional variance of the GARCH process $\mathbb{E}[\sigma_t^2] = \gamma_0 / (1 - \gamma_1 - \gamma_2)$ between $0.5$ and $0.01$ by varying the parameter $\gamma_0$ while holding $\gamma_1$ and $\gamma_2$ constant.
	Since the conditional variance is a weighted combination of the unconditional variance, the past squared returns and the past conditional variance, this change implies that the ES forecasts are too conservative when the unconditional variance is larger than its true value, and vice versa.
	
	\item
	We vary the persistence of shocks between $0.9$ and $0.999$ by setting $\tilde{\gamma}_1 = c\cdot\gamma_1$ and $\tilde{\gamma}_2 = c\cdot\gamma_2$ for a varying constant $c$ and by setting $\tilde{\gamma}_0 = \mathbb{E} [\sigma_t^2] (1-\tilde{\gamma}_1 - \tilde{\gamma}_2)$ in order to stabilize the unconditional variance.
	A higher persistence causes a stronger and longer reaction to shocks.
	
	\item
	We vary the degrees of freedom of the underlying Student-$t$ distribution between 3 and $\infty$.
	Since the conditional variance is unaffected, this modification implies a relative horizontal shift of the ES forecasts.
	
	\item
	We misspecify the probability level $\tilde{\tau}$ of the ES forecasts between $0.5\%$ and $5\%$.
	This represents the scenario that a forecaster submits (accidentally or on purpose) predictions for some level $\tilde{\tau} \neq \tau$.
	Similar to changing the degrees of freedom, this modification implies a relative horizontal shift of the ES forecasts.
\end{enumerate}
As an illustrative example of these misspecifications, \Cref{fig:mc2_example_series_1,fig:mc2_example_series_2,fig:mc2_example_series_3,fig:mc2_example_series_4,fig:mc2_example_series_5} in Appendix \ref{sec:AdditionalMaterial} depict 250 realizations of the returns of the true DGP in (\ref{eq:mc2_model}), together with the corresponding ES forecasts of the true model (black dashed line) and of two exemplary models following the parameter misspecifications described in the points (a) to (e) above.

\begin{figure}
	\centering
	\begin{subfigure}{.49\linewidth}
		\caption{Changing the ARCH parameter}
		\label{fig:mc2_rejection_rates_1}
		\includegraphics{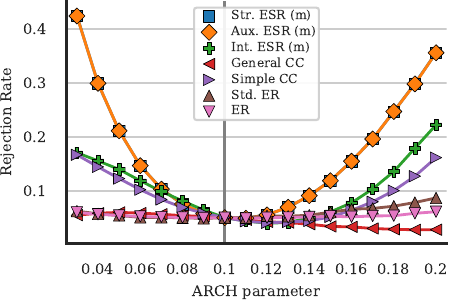}
	\end{subfigure}
	\hfill
	\begin{subfigure}{.49\linewidth}
		\caption{Changing the unconditional variance}
		\label{fig:mc2_rejection_rates_2}        
		\includegraphics{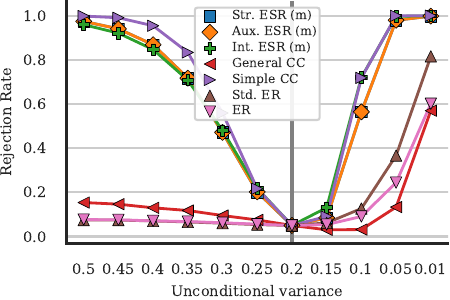}
	\end{subfigure}
	\\[\baselineskip]
	\begin{subfigure}{.49\linewidth}
		\caption{Changing the persistence}    
		\label{fig:mc2_rejection_rates_3}
		\includegraphics{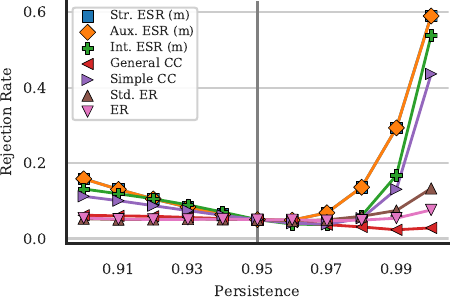}
	\end{subfigure}
	\\[\baselineskip]
	\begin{subfigure}{.49\linewidth}
		\caption{Changing the degrees of freedom}        
		\label{fig:mc2_rejection_rates_4}
		\includegraphics{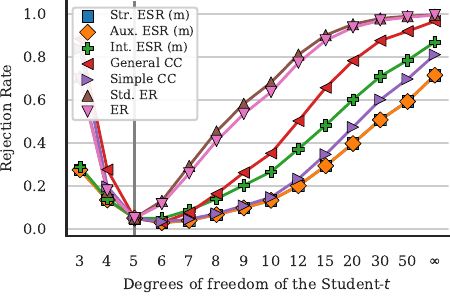}
	\end{subfigure}	
	\hfill
	\begin{subfigure}{.49\linewidth}
		\caption{Changing the probability level}        
		\label{fig:mc2_rejection_rates_5}
		\includegraphics{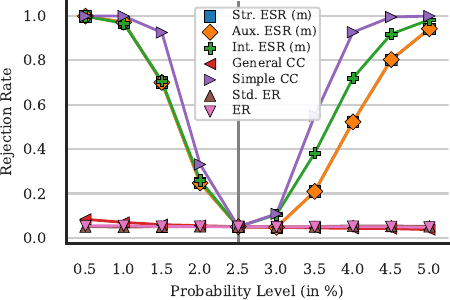}
	\end{subfigure}	    
	\caption[Power Plots Continuous Misspecification]{Size-adjusted rejection rates for various types of misspecification. The gray vertical line depicts the true model. 
	The number of Monte-Carlo repetitions is 10,000 and the probability level for the risk measures is $\tau=2.5\%$. ESR refers to the backtests introduced in this paper with (m) indicating the version which account for the additional covariance terms induced by the misspecified model. CC refers to the conditional calibration tests of \citet{Nolde2017}, and ER to the exceedance residuals tests of \citet{McNeil2000}.
	}
	\label{fig:mc2_rejection_rates}    
\end{figure}

We present the size-adjusted rejection rates plotted against the respective misspecified parameters for these five designs in \Cref{fig:mc2_rejection_rates_1,fig:mc2_rejection_rates_2,fig:mc2_rejection_rates_3,fig:mc2_rejection_rates_4,fig:mc2_rejection_rates_5}.
The true model is indicated by the gray vertical line and, induced by the results of \Cref{fig:mc2_example_series} in Appendix \ref{sec:AdditionalMaterial}, the x-axis is oriented such that too risky (too small in absolute value) ES forecasts are on the right side of the true model.\footnote{Notice that this inequality of the forecast magnitude only holds on average in the cases of \Cref{fig:mc2_rejection_rates_1,fig:mc2_rejection_rates_3} whereas it holds strictly for \Cref{fig:mc2_rejection_rates_2,fig:mc2_rejection_rates_4,fig:mc2_rejection_rates_5}.} 
Even though there is no backtest that dominates the others throughout all considered designs, several conclusions can be drawn from this figure.

\noindent
(1) Overall, the Strict and Auxiliary ESR tests perform almost indistinguishable and in four out of the five considered designs, their performance is superior compared to the general CC and both ER backtesting approaches. (\Cref{fig:mc2_rejection_rates_1,fig:mc2_rejection_rates_2,fig:mc2_rejection_rates_3,fig:mc2_rejection_rates_5}). 
The ESR backtests outperform the competitors especially when we misspecify the volatility dynamics of the underlying GARCH process (\Cref{fig:mc2_rejection_rates_1,fig:mc2_rejection_rates_2,fig:mc2_rejection_rates_3}).
This shows that, in contrast to the existing approaches, our ESR backtests can be used to detect misspecifications in the dynamics used to construct the ES forecasts which go beyond level shifts.

\noindent
(2) The two ER tests (and the general CC test that is constructed to be similar to the ER backtest) can hardly discriminate between forecasts for the VaR and ES issued through misspecified volatility processes  (\Cref{fig:mc2_rejection_rates_1,fig:mc2_rejection_rates_2,fig:mc2_rejection_rates_3}) and through misspecified probability levels $\tilde{\tau} \neq \tau$ (\Cref{fig:mc2_rejection_rates_5}).
This confirms the theoretical results discussed in Section \ref{sec:er_test} in Appendix \ref{sec:existing_backtests} that these backtests only reject misspecifications which affect the relation (distance) between the VaR and ES forecasts.
In contrast, these backtests perform well in the case of misspecified tails of the residual distribution, which particularly affects the relative distance between the VaR and ES forecasts (\Cref{fig:mc2_rejection_rates_4}).
If these backtests would be used by the regulatory authorities, banks could submit joint VaR and ES forecasts for some level $\tilde{\tau} > \tau$ or some (too small) volatility process in order to minimize their capital requirements without facing the risk of being detected by these backtests.
In comparison, our Intercept ESR backtest which is similar to the ER backtests by construction is clearly able to identify these misspecified probability levels. 

\noindent
(3) Throughout all five misspecifications, the simple CC backtest also exhibits good power properties, similar to our proposed backtests.
However, our three ESR backtests exhibit much better size properties (see \Cref{tab:mc1_size1} and \Cref{tab:mc2_size}) and in contrast to the simple CC test, they do not fail to reject the HS forecasts in the first simulation study (see \Cref{fig:MC1_power_plots}).

Together with the results from the first simulation study, these findings demonstrate that our proposed ESR backtests are a powerful choice for backtesting ES forecasts.
They are reasonably sized and exhibit good power properties against a variety of misspecifications.
Notably, in contrast to the existing backtests, there is no single type of misspecification where our ESR tests are unable to discriminate between forecasts of the true and the misspecified models.

\subsection{Testing One-Sided Hypotheses}
\label{sec:one_sided_alternative}

For the regulatory authorities, testing against a one-sided alternative might be more meaningful than the two-sided versions of the tests we consider in the previous sections.
Holding more money than stipulated bv the Basel Accords is no concern for regulators as it is only important that banks keep enough monetary reserves to cover the risks from their market activities.
In the following, we assess the performance of the Intercept ESR backtest and the one-sided versions of the four competitor backtests in rejecting the null hypothesis that the issued ES forecasts are at least as conservative (not smaller in absolute value) as the true ES, i.e. that the associated market risk is not underestimated.

\begin{figure}
	\centering
	\begin{subfigure}{.49\linewidth}
		\caption{Changing the reaction to the squared returns}
		\label{fig:mc2_rejection_rates_1_1s}
		\includegraphics{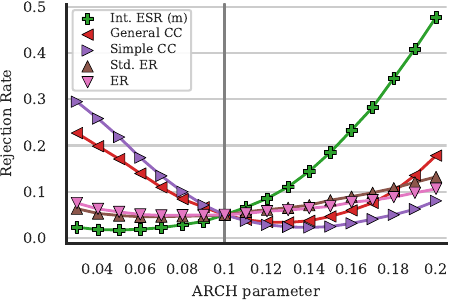}
	\end{subfigure}
	\hfill
	\begin{subfigure}{.49\linewidth}
		\caption{Changing the unconditional variance}
		\label{fig:mc2_rejection_rates_2_1s}  
		\includegraphics{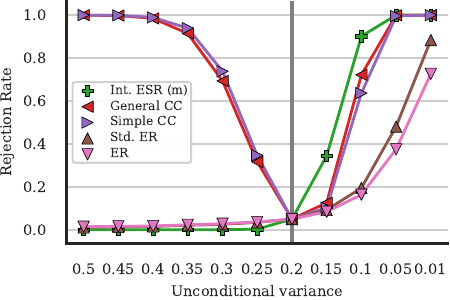}
	\end{subfigure}
	\\[\baselineskip]
	\begin{subfigure}{.49\linewidth}
		\caption{Changing the persistence}    
		\label{fig:mc2_rejection_rates_3_1s}
		\includegraphics{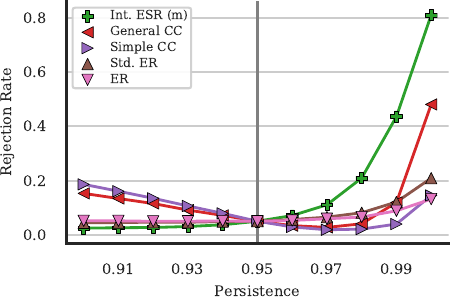}
	\end{subfigure}
	\\[\baselineskip]
	\begin{subfigure}{.49\linewidth}
		\caption{Changing the degrees of freedom}        
		\label{fig:mc2_rejection_rates_4_1s}
		\includegraphics{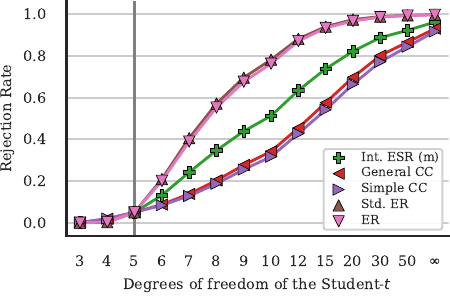}
	\end{subfigure}	
	\hfill
	\begin{subfigure}{.49\linewidth}
		\caption{Changing the probability level}        
		\label{fig:mc2_rejection_rates_5_1s}
		\includegraphics{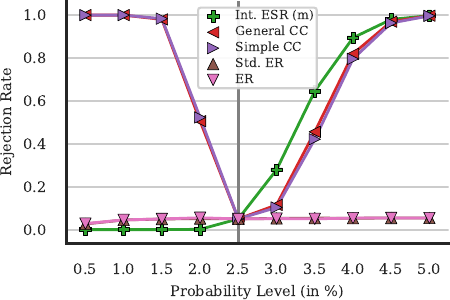}
	\end{subfigure}	
	\caption[Power Plots, One-Sides Hypothesis]{Size-adjusted rejection rates for various types of misspecification with a one-sided hypothesis. The gray vertical line depicts the true model. 
	The number of Monte-Carlo repetitions is 10,000 and the probability level for the risk measures is $\tau=2.5\%$. ESR refers to the backtests introduced in this paper with (m) indicating the version which account for the additional covariance terms induced by the misspecified model. CC refers to the conditional calibration tests of \citet{Nolde2017}, and ER to the exceedance residuals tests of \citet{McNeil2000}.
	}
	\label{fig:mc2_rejection_rates_1s}    
\end{figure}

In \Cref{fig:mc2_rejection_rates_1_1s,fig:mc2_rejection_rates_2_1s,fig:mc2_rejection_rates_3_1s,fig:mc2_rejection_rates_4_1s,fig:mc2_rejection_rates_5_1s}, we present the size-adjusted rejection rates for the one-sided versions of the considered backtests and for the five continuous parameter misspecifications described in the points (a) - (e) from the previous section.
The structure of these figures is analog to the two-sided case where the x-axis is oriented such that too risky ES forecasts are on the right side of the true model (vertical gray line).
As it can be seen in \Cref{fig:mc2_example_series_1,fig:mc2_example_series_2,fig:mc2_example_series_3,fig:mc2_example_series_4,fig:mc2_example_series_5} in Appendix \ref{sec:AdditionalMaterial}, the five modifications of the true model exhibit clear patterns when they issue too risky, respectively too conservative forecasts for the true ES, where this finding holds \textit{strictly} for the cases (b), (d) and (e) and \textit{on average} for the cases (a) and (c).
Thus, the one-sided backtests should only reject the null hypothesis for ES forecasts that issue too risky (too small in absolute value) forecasts, i.e. which are on the right side of the true model in \Cref{fig:mc2_rejection_rates_1_1s,fig:mc2_rejection_rates_2_1s,fig:mc2_rejection_rates_3_1s,fig:mc2_rejection_rates_4_1s,fig:mc2_rejection_rates_5_1s}.

We find that our Intercept ESR backtest is reasonably sized (compare \Cref{tab:mc2_size}) and clearly dominates the ER and the CC tests in terms of their power in four out of the five misspecification designs.
Only when altering the degrees of freedom, the ER tests are slightly more powerful than the Intercept ESR test.
Surprisingly, we see that in four out of the five cases, the one-sided CC tests (both, the simple and the general version) also reject too conservative ES forecasts, even though these should not be rejected by the specifications of the one-sided tests.
Furthermore, as for the two-sided tests, both ER backtests fail to detect misspecifications of the underlying volatility process and of the underlying probability level.
Summarizing these results, the proposed Intercept ESR backtest is a powerful backtest with good size properties for testing one-sided hypotheses which clearly dominates the existing one-sided (joint VaR and ES) backtesting techniques in the literature.

\section{Empirical Application}
\label{sec:empirical_application}

In the empirical application we apply our backtests to compare ES forecasts along three dimensions: the complexity of the risk model, the length of the estimation window, and the model refit frequency.
From a practitioners point of view, it would be desirable to have a parsimonious model that can be estimated with few observations and is valid over a long period of time, for reasons of low engineering effort, data storage, and human and computational effort for updating the model.
To assess whether such a setup is reasonable, and if not, which dimensions are crucial for a good performance, we compare rejection rates of ES forecasts using our backtests.

For this application, we use daily log returns of the 200 most highly capitalized stocks of the S\&P\,500 index (as of September 1, 2019), with a sufficiently long history of stock prices.
We consider four different risk models: the standard GARCH(1,1) of \citet{Bollerslev1986} and the GJR-GARCH(1,1) model of \cite{Glosten1993}, both coupled with Gaussian and Student-$t$ distributed innovations.
For all four models and 200 stocks, we compare the same evaluation horizon, the period from January 2010 to August 2019 with a total of 2432 daily observations.
We furthermore consider five different lengths of the rolling estimation window ranging from one year (250 trading days) up to eight years (2000 trading days) and refit horizons of 5, 21, 62, 125 and 250 days, corresponding to weekly, monthly, quarterly, bi-yearly and yearly updating of the models.

\Cref{tab:empirical_application} presents the rejection rates of the one-sided intercept ESR backtest with a nominal size of $5\%$ for the 200 stocks under investigation, for the four GARCH specifications, the five estimation window sizes and the five refit frequencies.
We choose to use the one-sided intercept test as this is the only one-sided and strict ES backtest in the literature.
Given the currently implemented traffic light system of the Basel Committee, this one-sided test might be the one with the highest practical relevance for backtesting the ES.


\begin{table}[ht]
	\centering	
	\caption{Results of the empirical application}
	\label{tab:empirical_application}
	\begin{tabularx}{\linewidth}{X *{5}{c} p{0.2cm} *{5}{c}}
		\toprule
		\multirow{2}{*}{\shortstack[l]{Rolling \\ Window}} & \multicolumn{5}{c}{Refit Frequency} && \multicolumn{5}{c}{Refit Frequency}\\
		\cmidrule(lr){2-6}	\cmidrule(lr){8-12}
		& 5 & 21 & 62 & 125 & 250 & & 5 & 21 & 62 & 125 & 250 \\
		\cmidrule(lr){2-6}	\cmidrule(lr){8-12}
		& \multicolumn{5}{c}{\underline{GARCH-$N$}} & & \multicolumn{5}{c}{\underline{GJR-GARCH-$N$}} \\
		 250 & 1.00 & 1.00 & 1.00 & 1.00 & 1.00 &  & 1.00 & 1.00 & 1.00 & 1.00 & 1.00 \\ 
  500 & 0.99 & 0.99 & 0.99 & 0.99 & 0.99 &  & 1.00 & 1.00 & 1.00 & 0.99 & 1.00 \\ 
  1000 & 0.98 & 0.97 & 0.98 & 0.96 & 0.96 &  & 0.99 & 0.99 & 0.99 & 0.99 & 0.99 \\ 
  1500 & 0.97 & 0.97 & 0.97 & 0.97 & 0.96 &  & 0.98 & 0.99 & 0.99 & 0.99 & 0.99 \\ 
  2000 & 0.96 & 0.96 & 0.94 & 0.95 & 0.94 &  & 0.98 & 0.98 & 0.98 & 0.98 & 0.97 \\ 
  
		\addlinespace
		& \multicolumn{5}{c}{\underline{GARCH-$t$}} & & \multicolumn{5}{c}{\underline{GJR-GARCH-$t$}} \\
		 250 & 0.26 & 0.32 & 0.32 & 0.36 & 0.59 &  & 0.28 & 0.32 & 0.29 & 0.37 & 0.42 \\ 
  500 & 0.10 & 0.09 & 0.12 & 0.12 & 0.20 &  & 0.13 & 0.17 & 0.17 & 0.22 & 0.25 \\ 
  1000 & 0.07 & 0.07 & 0.07 & 0.08 & 0.09 &  & 0.14 & 0.11 & 0.11 & 0.12 & 0.14 \\ 
  1500 & 0.09 & 0.09 & 0.09 & 0.10 & 0.09 &  & 0.08 & 0.10 & 0.09 & 0.09 & 0.09 \\ 
  2000 & 0.10 & 0.08 & 0.08 & 0.08 & 0.09 &  & 0.09 & 0.09 & 0.10 & 0.09 & 0.10 \\ 
  
		\bottomrule
		\addlinespace
		\multicolumn{12}{p{.97\linewidth}}{\footnotesize \textit{Notes:} This tables shows the rejection rates of the one-sided ESR backtest for ES forecasts stemming from the two GARCH-type models with Student's $t$ and Gaussian residuals, different rolling window sizes and model refit lengths (in days). 
		The rejection frequencies are averaged over the analyzed 200 most capitalized stocks of the S\&P\,500 index.
		The out-of-sample window covers the time from Jan. 2010 to Aug. 2019 resulting in a sample size of 2432 days.}
	\end{tabularx}
\end{table}

The results show that both, the GARCH-$N$ and GJR-GARCH-$N$ are rejected for almost all the stocks (in more than 94\% of the cases) uniformly over the different estimation sample sizes and refit frequencies.
Independent of the sample length and refit frequency, this supports the well known finding that Gaussian residuals generally fail to capture the riskiness of financial assets, especially in the tail of the distribution. 
In contrast, for the two GARCH specifications with Student-$t$ distributed innovations, the rejection frequencies are considerably lower and for many choices of refit frequencies and estimation windows, they are just above the nominal significance level.
These results imply that using fat-tailed distributions generally decreases the rejection frequency.
Similarly, refitting the models more frequently tends to decrease the rejection frequency, especially if the estimation window is short.
However, refitting the model on a monthly or even weekly basis is not required, infrequent regular updates (such as quarterly) suffice, as more frequent updates do not improve performance if the estimation window is large enough.
Interestingly, employing the GJR-GARCH model, which accounts for a potential leverage effect in the volatility, does not perform better than the standard GARCH model.

Generally, refitting these models at least quarterly using data which goes at least four years into the past suffices to obtain rejection rates uniformly below $11\%$.
The results of this application, which are diversified over 200 individual stocks, indicate that in order to obtain satisfactory ES forecasts, one should use fat-tailed residual distributions, more than four years of data and regular refitting of the models at least once per quarter.

\section{Conclusion}
\label{sec:conclusion}

With the upcoming implementation of the third Basel Accords, risk managers and regulators will shift attention to the risk measure Expected Shortfall (ES) for the forecasting and evaluation of financial risks.
In this paper, we introduce regression based \textit{ESR backtests} for ES forecasts, which extend the classical \cite{MincerZarnowitz1969} test to ES specific versions.
As estimation of regression parameters for the ES stand-alone is infeasible, our tests build on a recently developed joint VaR and ES regression, which allows for different specifications of our tests, titled the \textit{Auxiliary}, \textit{Strict}, and \textit{Intercept} ESR backtests.
As these tests are potentially subject to model misspecification, we extend the asymptotic theory for the joint VaR and ES regression framework to possibly misspecified models and verify the tests' performance in finite samples through an extensive simulation study.
We apply our tests to 200 stocks from the S\&P\,500 index in order to analyze the performance of ES forecasts stemming from the GARCH model family.
We find that using fat-tailed (Student's $t$) residual distributions, more than four years of data and regular refitting of the models at least once per quarter yields satisfactory ES forecasts.


A unique and essential feature of the Strict and Intercept ESR backtests is that they solely require forecasts for the ES and are consequently the first backtests for the ES stand-alone.
In contrast, a common drawback of the existing backtests in the literature is that they need forecasts of further input parameters, such as the VaR, the volatility, the tail distribution or even the whole return distribution.
Using more information than the ES forecasts is problematic for two reasons.
First, these tests are not applicable for the regulatory authorities, who receive forecasts of the ES, but not of the additional information required by these tests.
Second, rejecting the null hypothesis does not necessarily imply that the ES forecasts are incorrect as the rejection can be a result of a false prediction of any of the input parameters.

This paper contributes to the ongoing discussion about which risk measure is the best in practice in the following way.
As the VaR is criticized for not being subadditive and for not capturing tail risks beyond itself, the recent literature proposes both, the ES and expectiles as alternative risk measures.
Expectiles are suggested as they are coherent, elicitable and are able to capture extreme risks beyond the VaR and thus, they simultaneously overcome the drawbacks of the VaR and the ES \citep{Bellini2104, Ziegel2016}. Unfortunately, as opposed to the VaR and ES, they lack a visual and intuitive interpretation \citep{Emmer2015}.
In contrast, the ES is mainly criticized for its theoretical deficiencies of being not elicitable and not (only with difficulties) backtestable.
However, starting with the joint elicitability result of VaR and ES of \cite{Fissler2016}, there is a growing body of literature using this result for a regression procedure \citep{DimiBayer2019,Barendse2017,Patton2019} and for relative forecast comparison \citep{Fissler2016b,Nolde2017}, which is extended by this paper through introducing the ESR backtests, which are the first sensible backtests for the ES stand-alone.
This shows that, even though technically more demanding, the ES can be modeled, evaluated and backtested in the same way as quantiles and expectiles.
Combining this with its ability to capture extreme tail risks and its intuitive visual interpretation, the ES is an appropriate candidate for being the standard risk measure in practice.

\section*{Acknowledgments}
We thank the editor Andrew Patton, an anonymous associate editor and two referees for very helpful comments.
We further thank Tobias Fissler, Lyudmila Grigoryeva, Roxana Halbleib, Phillip Heiler, Ekaterina Kazak, Winfried Pohlmeier, James Taylor, and Johanna Ziegel for suggestions which inspired some results of this paper.
Financial support by the Heidelberg Academy of Sciences and Humanities (HAW) within the project ``Analyzing, Measuring and Forecasting Financial Risks by means of High-Frequency Data'' and by the German Research Foundation (DFG) within the research group ``Robust Risk Measures in Real Time Settings'' is gratefully acknowledged.
The authors acknowledge support by the state of Baden-Württemberg through bwHPC.
The majority of the work on this paper was conducted while both authors were at the Department of Economics, Universität Konstanz.

\FloatBarrier

\begin{appendices}

\renewcommand\thetable{\thesection.\arabic{table}}

\section{Proofs}
\label{sec:Proofs}

\begin{proof}[Proof of Theorem \ref{thm:ConsistencyMestimatorMisspecifiedModel}] 
	We check that the necessary conditions (i) - (iv) of the basic consistency theorem, given in Theorem 2.1 in \cite{NeweyMcFadden1994}, p.2121 hold, where we consider the objective functions $Q_{T}(\theta)$ and $Q_{T}^0(\theta)$ as defined in (\ref{eqn:QTObjectiveFunction}) and (\ref{eqn:DefinitionPseudoTrueParameter}).
	First, notice that condition (ii) holds by imposing condition \ref{cond:CompactParameterSpaceESReg}.
	The unique identification condition (i) holds by assumption \ref{cond:UniqueMinimum}.
	Next, we verify the uniform convergence condition (iv) by applying the uniform weak law of large numbers given in Theorem A.2.5. in \cite{White1994}. For that, we have to show that
	\begin{enumerate}[label=(\Alph*)]
		\item 
		\label{cond:LipschitzL1ESReg}
		the map $\theta \mapsto \rho \big( Y_t, X_t, \theta\big)$ is Lipschitz-$L_1$ on $\Theta$, see Definition A.2.3 in \cite{White1994}\footnote{
			Notice that we do not have a double index and thus we supress the $n$ in the notation of \cite{White1994}. Furthermore, we apply the definition by using the identify function for $a_t^{o}$.
		}, 
		
		\item 
		\label{cond:LLNlocally}
		For all $\theta^o \in \Theta$, there exists $\delta^o > 0$, such that for all $\delta, 0 < \delta \le \delta^o$, the sequences 
		\begin{align}
		\bar \rho_t(\theta^o,\delta) &:=  \sup_{\theta \in \Theta} \left\{ \left. \rho \big( Y_t, X_t, \theta \big) \right| || \theta - \theta^o || < \delta \right\} \qquad \text{and} \\
		\ubar{\rho}_t(\theta^o,\delta) &:=  \inf_{\theta \in \Theta} \left\{ \left. \rho \big( Y_t, X_t, \theta \big) \right| || \theta - \theta^o || < \delta \right\}
		\end{align}
		obey a weak law of large numbers.
	\end{enumerate}
	Condition \ref{cond:LipschitzL1ESReg} follows directly from Lemma \ref{lemma:LipschitzL1} and we turn to condition \ref{cond:LLNlocally}.
	As the process $\{Y_t, V_t, W_t\}$ is strong mixing of size $-r/(r-2)$ for some $r>2$ by condition \ref{cond:StrongMixingESReg}, the processes $V_t$ and $W_t$ are strong mixing of the same size by Theorem 3.49 in \cite{White2001}, p. 50.
	As the functions $\rho \big( Y_t, X_t, \theta \big)$ and the supremum/infimum functions are $\mathcal{F}_t$-measureable for all $t \in \mathbb{N}$, we can conclude that the sequences $\bar \rho_t(\theta^o,\delta)$ and $\ubar{\rho}_t(\theta^o,\delta)$ are also strong mixing of the same size by applying the same theorem.

	Furthermore, for $\tilde r > 1$ and for some $\delta > 0$ sufficiently small enough, $r \ge \tilde r+\delta$ and thus $\mathbb{E} \left[ |\bar{\rho}_t(\theta^o,\delta)|^{\tilde r + \delta} \right] \le \sup_{1\le t \le T}\mathbb{E} \left[ \sup_{\theta \in \Theta} \left| \rho \big( Y_t, X_t, \theta \big) \right|^{r} \right]$ for all $t, 1\le t \le T, T\ge1$.
	As $\Theta$ is compact, there exists some $c > 0$ such that $\sup_{\theta \in \Theta} || \theta|| \le c$ and thus, for all $t = 1,\dots,T$, it holds that
	\begin{align}
		\mathbb{E} \left[ \sup_{\theta \in \Theta} \left| \rho \big( Y_t, X_t, \theta \big) \right|^{r} \right]
		\le 4^{r-1} \left\{ 
		1 
		+ \left( \frac{c}{K} \left( 1+ \frac{1}{\tau} \right) \right) \mathbb{E} || V_t||^r
		+ \frac{1}{\tau K} \mathbb{E} || Y_t||^r
		+ \sup_{\theta \in \Theta} \mathbb{E} || \log(W_t^\top \gamma )||^r
		\right\},
	\end{align}
	which is bounded by condition \ref{cond:MomentConditionsESReg} and as $\log(z) \le z$ for $z$ large enough.
	The same inequality holds for $|\ubar{\rho}_t(\theta^o,\delta)|$.
	Thus, we can apply the weak law of large numbers for strong mixing sequences in Corollary 3.48 in \cite{White2001}, p. 49 in order to conclude that for all $\theta^o \in \Theta$ such that $||\theta^o - \theta || \le \delta$, it holds that $\frac{1}{T} \sum_{t=1}^T \big( \bar{\rho}_t(\theta^o,\delta) -  \mathbb{E} \left[ \bar{\rho}_t(\theta^o,\delta) \right] \big) \stackrel{\mathbb{P}}{\to} 0$ and $\frac{1}{T} \sum_{t=1}^T \big( \ubar{\rho}_t(\theta^o,\delta) - \mathbb{E} \left[ \ubar{\rho}_t(\theta^o,\delta) \right] \big) \stackrel{\mathbb{P}}{\to} 0$, which shows condition \ref{cond:LLNlocally}.
	Consequently, the uniform convergence condition (iv) holds by applying the uniform weak law of large numbers given in Theorem A.2.5. in \cite{White1994}.
	
	As we have shown that the map $\theta \mapsto \rho \big( Y_t, X_t, \theta \big)$ is Lipschitz-$L_1$  in Lemma \ref{lemma:LipschitzL1}, the map $\theta \mapsto Q_{T}^0 = \frac{1}{T} \sum_{t=1}^T \mathbb{E} \left[  \rho \big( Y_t,X_t,\theta \big) \right]$ is also continuous which shows condition (iii).
	Thus, we can apply Theorem 2.1. of \cite{NeweyMcFadden1994} which concludes the proof of this theorem.
\end{proof}


\begin{proof}[Proof of Theorem \ref{thm:AsymptoticNormalityMestimatorMisspecifiedModel}]
	Let
	\begin{align}
		\label{eqn::RegressionIdFunction}  
		\psi(Y_t, X_t, \theta) 
		= \begin{pmatrix}
		- \frac{V_t}{\tau W_t^\top \gamma} \big( \mathds{1}_{\{Y_t \le V_{t}^\top \beta \}} - \tau \big) \\
		\frac{W_t}{ (W_t^\top \gamma)^2} \left( W_{t}^\top \gamma  - V_{t}^\top \beta + \frac{1}{ \tau}  (V_{t}^\top \beta - Y_t) \mathds{1}_{\{Y_t \le V_{t}^\top \beta \}} \right)
		\end{pmatrix},
	\end{align}
	which is almost surely the derivative of $\rho(Y_t,X_t,\theta)$ with respect to $\theta$.
	We further define $\Psi_{T}(\theta) = \frac{1}{T} \sum_{t=1}^T \psi(Y_t,X_t,\theta)$ and $\Psi_{T}^0(\theta) = \mathbb{E} [ \Psi_{T}(\theta) ]$.
	From the proof of Lemma \ref{lemma:WeissConditionN3iESReg}, we get the mean value expansion (for $\hat \theta_T$ close to $\theta^\ast_T$),
	\begin{align}
		\label{eqn:ProofApplicationMeanValue}
		\Psi_{T}^0(\hat \theta_T) - \Psi_{T}^0(\theta_T^\ast) = \Delta_T(\tilde \theta_1, \tilde \theta_2) \big( \hat \theta_T - \theta^\ast_T \big),
	\end{align} 
	for some values $\tilde \theta_1$ and $\tilde \theta_2$ somewhere on the line between $\hat \theta_T$ and $\theta_T^\ast$, where the components of $\Delta_T(\tilde \theta_1,\tilde \theta_2)$ are given in (\ref{eqn:ComponentsDeltaMeanValue1}) and  (\ref{eqn:ComponentsDeltaMeanValue}), and where $\Psi_{T}^{0}(\theta_T^\ast) = 0$.\footnote{The mean-value theorem cannot be generalized in a straight-forward fashion to vector-valued functions. Thus, we have to consider the mean value expansion in each component separately which gives this more complicated expression.}
	
	Furthermore, it holds that $\Delta_T(\theta^\ast_T, \theta^\ast_T) = \Lambda_T(\theta^\ast_T) $ and $\Delta_T(\tilde \theta_1, \tilde \theta_2)$ is a continuous function in its arguments $\tilde \theta_1$ and $\tilde \theta_2$.
	Using that $\Lambda_T(\theta^\ast_T) $ has Eigenvalues bounded away from zero (for $T$ large enough), we also get that $\Delta_T(\tilde \theta_1, \tilde \theta_2)$ is non-singular in a neighborhood around $\theta^\ast_T$ (for all arguments) for $T$ large enough as the map which maps the matrix onto its Eigenvalues is continuous.
	As we further know that $\hat \theta_T -  \theta^\ast_T \stackrel{\mathbb{P}}{\to} 0$ and $|| \tilde \theta_j - \theta_T^\ast || \le || \hat \theta_T - \theta^\ast_T ||$ for all $j = 1,2$, we get from the continuous mapping theorem that
	\begin{align}
	\label{eqn:ContinuousMappingDeltaT}
	\Delta_T^{-1}(\tilde \theta_1, \tilde \theta_2) - \Lambda_T^{-1}(\theta^\ast_T)  \stackrel{\mathbb{P}}{\to} 0.
	\end{align}

	In the following, we apply Lemma A.1 in \cite{Weiss1991} (by verifying its assumptions), which extends the iid results of \cite{Huber1967} to strong mixing sequences. 
	Assumption (N1) of Lemma A.1 in \cite{Weiss1991} is satisfied as every almost surely continuous stochastic process is separable in the sense of Doob \citep{GikhmanSkorokhod2004} and the functions $\psi \big( Y_t, X_t, \theta \big)$ are almost surely continuous for all $t \in \mathbb{N}$.
	Assumption (N2) is satisfied as shown in the proof of Theorem \ref{thm:ConsistencyMestimatorMisspecifiedModel}.
	Assumption (N3)(i) is shown in Lemma \ref{lemma:WeissConditionN3iESReg}.
	The technical Assumptions (N3)(ii) and (N3)(iii) follow from Lemma 4 and Lemma 5 in \cite{Patton2019Supplement}. 
	For this, notice that the moment conditions in Assumption 2 (C) and (D) of \cite{Patton2019} are implied by the condition \ref{cond:MomentConditionsESReg} in Assumption \ref{ass:Assumption1ESReg} for the simplified case of linear models.
	Assumption (N4) follows from the moment conditions \ref{cond:MomentConditionsESReg} in Assumption \ref{ass:Assumption1ESReg}  and Assumption (N5) from the strong mixing condition \ref{cond:StrongMixingESReg}. 
	Furthermore,  Lemma 2 of \cite{Patton2019Supplement} implies that $\sqrt{T} \Psi_{T}(\hat \theta_T) \stackrel{\mathbb{P}}{\to} 0$.
	Thus, we can apply Lemma A.1 in \cite{Weiss1991} and get that 
	\begin{align}
	\label{eqn:ResultWeiss1991ESReg}
	\sqrt{T} \Psi_{T}^0(\hat \theta_T) - \sqrt{T} \Psi_{T}(\theta^\ast_T) \stackrel{\mathbb{P}}{\to} 0.
	\end{align}
	Combining (\ref{eqn:ProofApplicationMeanValue}), (\ref{eqn:ContinuousMappingDeltaT}) and (\ref{eqn:ResultWeiss1991ESReg}), we get that
	\begin{align}
	\sqrt{T} \big( \hat \theta_T - \theta^\ast_T \big) &= - \Delta_T(\tilde \theta_1, \tilde \theta_2)^{-1} \sqrt{T} \, \Psi_T^0(\hat \theta_T) \\
	&= - \left( \Lambda_T^{-1}(\theta^\ast_T)  + o_p(1) \right) \cdot \left( \sqrt{T} \Psi_T(\theta^\ast_T) + o_p(1)\right)\\
	&= - \Lambda_T^{-1}(\theta^\ast_T)  \cdot \sqrt{T} \Psi_T(\theta^\ast_T) + o_p(1).
	\end{align}
	Furthermore, 
	\begin{align}
		\Sigma_T^{-1/2}(\theta^\ast_T) \sqrt{T} \Psi_T(\theta^\ast_T) = \Sigma_T^{-1/2}(\theta^\ast_T) \sqrt{T} \left( \Psi_T(\theta^\ast_T) - \Psi_T^0(\theta^\ast_T) \right) \stackrel{d}{\to} \mathcal{N} \big( 0, I_{2k} \big),
	\end{align}
	by Lemma \ref{lemma:AsymptoticNormalityPsiESReg} and thus,
	\begin{align}
	\Sigma_T^{-1/2}(\theta^\ast_T) \Lambda_T(\theta^\ast_T) \, \sqrt{T} \big( \hat \theta_T - \theta^\ast_T \big) \stackrel{d}{\to} \mathcal{N} \big( 0, I_{2k} \big),
	\end{align}		
	which concludes the proof of this theorem.
\end{proof}

\begin{proof}[Proof of Corollary \ref{cor:TestStatisticChiSq}]
	We first notice that
	\begin{align}
		\widehat \Omega_T^{-1/2}  \sqrt{T} \big( \hat \theta_T - \theta_T^\ast \big)
		=  \Omega_T^{-1/2}  \sqrt{T} \big( \hat \theta_T - \theta_T^\ast \big)
		+  \big( \widehat \Omega_T^{-1/2} - \Omega_T^{-1/2} \big)   \sqrt{T} \big( \hat \theta_T - \theta_T^\ast \big).
	\end{align}
	From Theorem \ref{thm:AsymptoticNormalityMestimatorMisspecifiedModel}, we get that  $\Omega_T^{-1/2}  \sqrt{T} \big( \hat \theta_T - \theta_T^\ast \big) \stackrel{d}{\to} \mathcal{N} \big( 0, I_4 \big)$.
	Furthermore, as $\big( \widehat \Omega_T^{-1/2} - \Omega_T^{-1/2} \big) = o_P(1)$ it holds by Slutzky's theorem, that $\big( \widehat \Omega_T^{-1/2} - \Omega_T^{-1/2} \big) \sqrt{T} \big( \hat \theta_T - \theta_T^\ast \big) = o_P(1)$ and consequently,
	\begin{align}
		\widehat \Omega_T^{-1/2}  \sqrt{T} \big( \hat \theta_T - \theta_T^\ast \big) \stackrel{d}{\to} \mathcal{N} \left( 0, I_4 \right).
	\end{align} 
	Thus,
	\begin{align}
		T_{\text{A-ESR}} &= \left( \widehat \Omega_{T,\gamma}^{-1/2}  \sqrt{T} \big( \hat \gamma_T - \gamma_T^\ast \big)  \right)^\top \left( \widehat \Omega_{T,\gamma}^{-1/2}  \sqrt{T} \big( \hat \gamma_T - \gamma_T^\ast \big)  \right) \stackrel{d}{\to} \chi^2_2, \\
		\tilde T_{\text{J-ESR}} &= \left( \widehat \Omega_{T,\gamma}^{-1/2}  \sqrt{T} \big( \hat \gamma_T - \gamma_T^\ast \big)  \right)^\top \left( \widehat 	\Omega_{T,\gamma}^{-1/2}  \sqrt{T} \big( \hat \gamma_T - \gamma_T^\ast \big)  \right) \stackrel{d}{\to} \chi^2_2, \qquad \text{ and } \\
		\tilde T_{\text{I-ESR}} &= \left( \widehat \Omega_{T,\gamma_1}^{-1/2}  \sqrt{T} \big( \hat \gamma_{T,1} - \gamma_{T,1}^\ast \big)  \right)^\top \left( \widehat \Omega_{T,\gamma_1}^{-1/2}  \sqrt{T} \big( \hat \gamma_{T,1} - \gamma_{T,1}^\ast \big)  \right) \stackrel{d}{\to} \chi^2_1.
	\end{align}
\end{proof}
	
\end{appendices}

\newpage
\begin{center}
{\Large\bf SUPPLEMENTARY MATERIAL}
\end{center}

\begin{appendices}

\section{Technical Proofs}
\label{sec:TechnicalProofs}

\begin{lemma}
	\label{lemma:LipschitzL1}
	Given the conditions from Assumption \ref{ass:Assumption1ESReg}, the function $\rho \big( Y_t, X_t, \theta \big)$ is $L_1$-Lipschitz on $\Theta$ with $\mathcal{F}_t$-measurable and integrable Lipschitz-constant.  
\end{lemma}

\begin{proof}
	We split the $\rho$-function $\rho  \big( Y_t, X_t, \theta \big) = \rho_1  \big( Y_t, X_t, \theta \big) + \rho_2  \big( Y_t, X_t, \theta \big)$, where
	\begin{align*} 
	\rho_1  \big( Y_t, X_t, \theta \big) &= - \mathds{1}_{\{Y_t \le V_t^\top \beta \}}  \frac{1}{\tau W_t^\top \gamma} ( V_t^\top \beta - Y_t) , \\
	\rho_2  \big( Y_t, X_t, \theta \big) &=  \frac{V_t^\top \beta - W_t^\top \gamma }{W_t^\top \gamma}  - \log(-W_t^\top \gamma).
	\end{align*}	
	Local Lipschitz continuity of $\rho_2$ follows since it is a continuously differentiable function in $\theta$ (such that $W_t^\top \gamma \not= 0$) and thus (locally) Lipschitz-$L_1$.
	We consequently get that for all $\theta^o \in \Theta$, there exists a $\delta^o > 0$ such that for all $\theta \in U_{\delta^o}(\theta^o) := \big\{ \theta \in \Theta \big| || \theta - \theta^o|| \le \delta^o \big\}$, it holds that
	\begin{align}
	\big| \rho_2 \big( Y_t, X_t, \theta^o \big)  - \rho_2 \big( Y_t, X_t, \theta \big)  \big| 
	\le \big|\big| \theta - \theta^o \big|\big|
	\cdot \sup_{\theta \in U_{\delta^o}(\theta^o)} \left( \left| \left| \frac{V_t + W_t}{W_t^\top \gamma} \right|\right| + \left| \left| \frac{ V_t^\top \beta W_t }{(W_t^\top \gamma)^2} \right|\right| \right),
	\end{align}
	where the sequences $\frac{1}{T} \sum_{t=1}^T \mathbb{E} \left[  \left| \left| \frac{V_t + W_t}{W_t^\top \gamma} \right|\right|\right]$ and $\frac{1}{T} \sum_{t=1}^T \mathbb{E} \left[ \left| \left| \frac{ V_t^\top \beta  W_t }{(W_t^\top \gamma)^2} \right|\right| \right]$ are bounded for all $\theta^o \in \Theta$ by the conditions \ref{cond:ESModelBounded} and \ref{cond:MomentConditionsESReg} in Assumption \ref{ass:Assumption1ESReg}.
	
	For the function $\rho_1$, we consider four cases.
	First, let $\Gamma_1 = \big\{ \omega \in \Omega, \theta \in U_{\delta^o}(\theta^o) \, \big| \, V_t^\top(\omega) \beta^o < Y_t(\omega) \; \text{ and } \; V_t^\top(\omega) \beta < Y_t(\omega) \big\}$.
	Then, on $\Gamma_1$, it holds that,
	\begin{align}
	\rho_1 \big( Y_t, X_t, \theta \big)  = \rho_1 \big( Y_t, X_t, \theta^o \big)  = 0,
	\end{align}
	which is obviously Lipschitz-$L_1$. 
	
	Second, let $\Gamma_2 = \big\{ \omega \in \Omega, \theta \in U_{\delta^o}(\theta^o) \, \big| \, V_t^\top(\omega) \beta^o \ge Y_t(\omega) \; \text{ and } \; V_t^\top(\omega) \beta \ge Y_t(\omega) \big\}$.
	On $\Gamma_2$, for both $\tilde \theta \in \{ \theta, \theta^o\}$, it holds that
	\begin{align}
	\rho_1 \big( Y_t, X_t, \tilde \theta \big)  = - \frac{1}{\tau W_t^\top \tilde \gamma}  \big( V_t^\top \tilde \beta - Y_t \big),
	\end{align}
	which is a continuously differentiable function. 
	Thus,
	\begin{align}
	\big|\rho_1 \big( Y_t, X_t \theta^o \big) - \rho_1 \big( Y_t, X_t ,\theta \big) \big|
	\le \big|\big| \theta^o - \theta \big|\big|
	\cdot  \left( \sup_{\theta \in U_{\delta^o}(\theta^o)} \left|\left|  \frac{ V_t }{\tau (W_t^\top \gamma)}   \right|\right|  +  \sup_{\theta \in U_{\delta^o}(\theta^o)}  \left|\left|  \frac{ W_t }{\tau (W_t^\top \gamma)^2} (V_t^\top \beta - Y_t)   \right|\right| \right),
	\end{align}
	where the average of the expectations of the suprema sequencesin the last two lines are bounded by the conditions \ref{cond:ESModelBounded} and \ref{cond:MomentConditionsESReg}  in Assumption \ref{ass:Assumption1ESReg}.
	
	Finally, let $\Gamma_3 = \big\{ \omega \in \Omega, \theta \in U_{\delta^o}(\theta^o)  \, \big| \,  V_t^\top(\omega) \beta < Y_t(\omega) \le V_t^\top(\omega) \beta^o \big\}$.
	As on $\Gamma_3$, $|V_t^\top \beta^o - Y_t | \le | V_t^\top \beta^o - V_t^\top \beta | $ almost surely, it holds that
	\begin{align*}
	&\big|\rho_1 \big( Y_t, X_t \theta^o \big) - \rho_1 \big( Y_t, X_t ,\theta \big) \big| 
	= \left| \frac{1}{\tau W_t^\top \gamma^o} ( V_t^\top \beta^o - Y_t) \right| \\
	\le \,&\left| \frac{1}{\tau W_t^\top \gamma^o} ( V_t^\top \beta^o - V_t^\top \beta) \right| 
	\le \, \big|\big| \theta - \theta^o \big|\big|
	\cdot \sup_{\theta \in U_{\delta^o}(\theta^o)} \left| \left| \frac{V_t}{\tau W_t^\top \gamma} \right|\right|.
	\end{align*}
	Equivalently as above, the average of the expectations of the suprema sequences in the last two lines are bounded by the condition \ref{cond:ESModelBounded} and \ref{cond:MomentConditionsESReg} in \ref{ass:Assumption1ESReg}.
	An equivalent argument holds for $\Gamma_4 = \big\{ \omega \in \Omega, \theta \in U_{\delta^o}(\theta^o)  \, \big| \,  V_t^\top(\omega) \beta^o < Y_t(\omega) \le  V_t^\top(\omega) \beta \big\}$.
	As $\Omega = \bigcup_{i=1}^4 \Gamma_i$, we can conclude that the function $\rho \big( Y_t, X_t, \theta \big)$ is Lipschitz-$L_1$ on $\Theta$.
\end{proof}

\begin{lemma}
	\label{lemma:WeissConditionN3iESReg}
	Given the conditions from Assumption \ref{ass:Assumption1ESReg}, there exist constants $a, d_0 > 0$ such that
	\begin{align}
	\big| \big| \Psi_{T}^0(\theta) \big| \big| \ge a || \theta - \theta_T^\ast || \qquad \text{ for any }  \theta \in \Theta \text{ such that }  ||\theta - \theta_T^\ast || \le d_0,
	\end{align}
	and for all $T \ge T_0$, where $T_0 \in \mathbb{N}$ is large enough.
\end{lemma}

\begin{proof}
	Let $\theta \in \Theta$ such that $||\theta - \theta_T^\ast || \le d_0$ for some (small) constant $d_0 > 0$ and define
	\begin{align}
	\Psi^0_{T,1}(\theta) &= \mathbb{E} \left[ - \frac{V_t}{\tau W_t^\top \gamma} \big( F_t(V_t^\top \beta) - \tau \big) \right] \qquad \text{ and } \\
	\Psi^0_{T,2}(\theta) &=  \mathbb{E} \left[ \frac{W_t}{(W_t^\top \gamma)^2} \left( W_{t}^\top \gamma  - V_{t}^\top \beta + \frac{1}{ \tau}  (V_{t}^\top \beta - Y_t) \mathds{1}_{\{Y_t \le V_{t}^\top \beta \}} \right) \right],
	\end{align}
	such that $\Psi^0_{T}(\theta)^\top  = \left( 	\Psi^0_{T,1}(\theta)^\top, \Psi^0_{T,2} (\theta)^\top  \right)$.
	Then, by applying the mean-value theorem we get that
	\begin{align}
	\begin{aligned}
	\Psi^0_{T,1}(\theta) - 	\Psi^0_{T,1}(\theta^\ast_T) 
	&= \mathbb{E} \left[ - \frac{V_t}{\tau W_t^\top \gamma} \big( F_t(V_t^\top \beta) - \tau \big) \right] - 
	\mathbb{E} \left[ - \frac{V_t}{\tau W_t^\top \gamma^\ast_T} \big( F_t(V_t^\top \beta^\ast_T) - \tau \big) \right] \\
	&=\mathbb{E} \left[ 
	\begin{pmatrix}
	- V_t V_t^\top \frac{1}{\tau W_t^\top \tilde \gamma_1}  f_t(V_t^\top \tilde \beta_1) \\
	V_t W_t^\top \frac{1}{\tau (W_t^\top \tilde \gamma_1)^2} \big( F_t(V_t^\top \tilde \beta_1) - \tau \big)
	\end{pmatrix}  \right]  \cdot (\theta- \theta_T^\ast) \\
	&= \Delta_{T,1} (\tilde \theta_1) \cdot (\theta- \theta_T^\ast),
	\label{eqn:ComponentsDeltaMeanValue1}
	\end{aligned}
	\end{align}
	for some $\tilde \theta_1$ on the line between $\theta$ and $\theta_T^\ast$.
	Equivalently, for the second component,
	\begin{align}
	\begin{aligned}
	&\Psi^0_{T,2}(\theta) - 	\Psi^0_{T,2}(\theta^\ast_T)  \\
	= \,&\mathbb{E} \left[ \frac{W_t}{(W_t^\top \gamma)^2} \left( W_{t}^\top \gamma  - V_{t}^\top \beta + \frac{1}{ \tau}  (V_{t}^\top \beta - Y_t) \mathds{1}_{\{Y_t \le V_{t}^\top \beta \}} \right) \right] \\
	&\qquad- \mathbb{E} \left[ \frac{W_t}{(W_t^\top \gamma^\ast_T)^2} \left( W_{t}^\top \gamma^\ast_T  - V_{t}^\top \beta^\ast_T + \frac{1}{ \tau}  (V_{t}^\top \beta^\ast_T - Y_t) \mathds{1}_{\{Y_t \le V_{t}^\top \beta^\ast_T \}} \right) \right] \\
	=\,&\mathbb{E} \left[ 
	\begin{pmatrix}
	W_t V_t^\top \frac{1}{(W_t^\top \tilde \gamma_2)^2} \frac{F_t(V_t \tilde \beta_2) - \tau }{\tau} \\
	W_t W_t^\top \left( \frac{1}{(W_t^\top \tilde \gamma_2)^2} -  \frac{2}{(W_t^\top \tilde \gamma_2)^3}  \left( W_{t}^\top \tilde \gamma_2  - V_{t}^\top \tilde \beta_2 + \frac{1}{ \tau}  (V_{t}^\top \tilde \beta_2 - Y_t) \mathds{1}_{\{Y_t \le V_{t}^\top \tilde \beta_2 \}} \right) \right)
	\end{pmatrix}  \right]  \cdot (\theta- \theta_T^\ast) \\
	= \, & \Delta_{T,2} (\tilde \theta_2) \cdot (\theta- \theta_T^\ast),
	\label{eqn:ComponentsDeltaMeanValue}
	\end{aligned}
	\end{align}
	for some $\tilde \theta_2$ on the line between $\theta$ and $\theta_T^\ast$.
	Notice that $\tilde \theta_1$ and $\tilde \theta_2$ are not necessarily the same as the mean-value theorem does not hold in its classical form for vector-valued functions.
	Thus, for $\Delta_T(\tilde \theta_1, \tilde \theta_2) =  \big( \Delta_{T,1}(\tilde \theta_1), \Delta_{T,2}(\tilde \theta_2) \big)$, we get that
	\begin{align}
	\Psi^0_{T}(\theta) - \Psi^0_{T}(\theta^\ast_T) = \Delta_T(\tilde \theta_1, \tilde \theta_2) \cdot (\theta -\theta^\ast_T).
	\end{align}
	In the following, we show that $\left| \left| \Delta_T \big(\tilde \theta_1, \tilde \theta_2 \big) - \Lambda_T(\theta^\ast_T)  \right| \right| \le c_1 || \theta - \theta^\ast_T ||$.
	For the first component of $\Delta_{T,1}( \tilde \theta_1)$, we get that
	\begin{align}
	|| \Delta_{T,11} (\tilde \theta_1)  - \Lambda_{T,11}(\theta^\ast_T)  ||
	&= \left| \left| \mathbb{E} \left[ 
	\begin{pmatrix}
	- V_t V_t^\top \frac{f_t'(V_t^\top \beta^{\ast \ast})}{\tau W_t^\top \gamma^{\ast \ast}}  V_t^\top (\tilde \beta_1 - \beta_T^\ast) \\
	V_t V_t^\top \frac{f_t(V_t^\top \beta^{\ast \ast})}{\tau (W_t^\top \gamma^{\ast \ast})^2}  W_t^\top (\tilde \gamma_1 - \gamma_T^\ast)
	\end{pmatrix}  \right] \right| \right| \\
	&\le \mathbb{E} \left[ 
	\begin{pmatrix}
	|| V_t||^3 \left| \frac{f_t'(V_t^\top \beta^{\ast \ast})}{\tau W_t^\top \gamma^{\ast \ast}} \right|  \\
	|| V_t ||^2 \cdot ||W_t|| \left| \frac{f_t(V_t^\top \beta^{\ast \ast})}{\tau (W_t^\top \gamma^{\ast \ast})^2}  \right|
	\end{pmatrix}  \right] \cdot || \tilde \theta_1 - \theta_T^\ast ||,
	\end{align}
	for some $\theta^{\ast \ast} = ( \beta^{\ast \ast},  \gamma^{\ast \ast})$ on the line between $\tilde \theta_1$ and $\theta_T^\ast$.
	
	For the second component of $\Delta_{T,1}( \tilde \theta_1)$ (and equivalently for the first component of of $\Delta_{T,2}( \tilde \theta_2)$), we get that
	\begin{align}
	|| \Delta_{T,12} (\tilde \theta_1)  - \Lambda_{T,12}(\theta^\ast_T)  ||
	&= \left| \left|  \mathbb{E} \left[ 
	\begin{pmatrix}
	- V_t W_t^\top \frac{f_t(V_t^\top \beta^{\ast \ast})}{\tau (W_t^\top \gamma^{\ast \ast})^2}  V_t^\top (\tilde \beta_1 - \beta_T^\ast) \\
	V_t W_t^\top \frac{2 (\tau - F_t(V_t^\top \beta^{\ast \ast}))}{\tau (W_t^\top \gamma^{\ast \ast})^3}  W_t^\top (\tilde \gamma_1 - \gamma_T^\ast)
	\end{pmatrix}  \right] \right| \right| \\
	&\le \mathbb{E} \left[ 
	\begin{pmatrix}
	|| V_t ||^2 \cdot ||W_t|| \left| \frac{f_t(V_t^\top \beta^{\ast \ast})}{\tau (W_t^\top \gamma^{\ast \ast})^2} \right|  \\
	|| V_t || \cdot ||W_t||^2 \left| \frac{2 (\tau - F_t(V_t^\top \beta^{\ast \ast}))}{\tau (W_t^\top \gamma^{\ast \ast})^3} \right| 
	\end{pmatrix}  \right]  \cdot || \tilde \theta_1 - \theta_T^\ast ||,
	\end{align}
	for some $\theta^{\ast \ast} = ( \beta^{\ast \ast},  \gamma^{\ast \ast})$ on the line between $\tilde \theta_1$ and $\theta_T^\ast$.
	
	Eventually, for the second component of $\Delta_{T,2}( \tilde \theta_2)$, we get that
	\begin{align}
	&|| \Delta_{T,22} (\tilde \theta_1)  - \Lambda_{T,22}(\theta^\ast_T)  || \\
	= \, &\left| \left|  \mathbb{E} \left[ 
	\begin{pmatrix}
	- W_t W_t^\top \frac{ 2 ( \tau - F_t(V_t^\top \beta^{\ast \ast}) ) }{\tau (W_t^\top \gamma^{\ast \ast})^3}  V_t^\top (\tilde \beta_1 - \beta_T^\ast) \\
	W_t W_t^\top  \left\{ \frac{-4}{(W_t^\top \gamma^{\ast \ast})^3} + \frac{6}{(W_t^\top \gamma^{\ast \ast})^4}  \left( W_{t}^\top  \gamma^{\ast \ast}  - V_{t}^\top \beta^{\ast \ast} + \frac{1}{ \tau}  (V_{t}^\top  \beta^{\ast \ast} - Y_t) \mathds{1}_{\{Y_t \le V_{t}^\top  \beta^{\ast \ast} \}} \right)  W_t^\top (\tilde \gamma_2 - \gamma_T^\ast) \right\}
	\end{pmatrix}  \right] \right| \right| \\
	\le \, &\mathbb{E} \left[ 
	\begin{pmatrix}
	|| V_t || \cdot ||W_t||^2 \left| \frac{ 2 ( \tau - F_t(V_t^\top \beta^{\ast \ast}) ) }{\tau (W_t^\top \gamma^{\ast \ast})^3}  \right|  \\
	||W_t||^3 \left|\frac{-4}{(W_t^\top \gamma^{\ast \ast})^3} + \frac{6}{(W_t^\top \gamma^{\ast \ast})^4}  \left( W_{t}^\top  \gamma^{\ast \ast}  - V_{t}^\top \beta^{\ast \ast} + \frac{1}{ \tau}  (V_{t}^\top  \beta^{\ast \ast} - Y_t) \mathds{1}_{\{Y_t \le V_{t}^\top  \beta^{\ast \ast} \}} \right) \right| 
	\end{pmatrix}  \right]  \cdot || \tilde \theta_2 - \theta_T^\ast ||,
	\end{align}
	for some $\theta^{\ast \ast} = ( \beta^{\ast \ast},  \gamma^{\ast \ast})$ on the line between $\tilde \theta_1$ and $\theta_T^\ast$.
	As the respective moments are finite given the moment conditions in \ref{cond:MomentConditionsESReg} in Assumption \ref{ass:Assumption1ESReg} and since $|| \tilde \theta_1 - \theta_T^\ast || \le || \theta - \theta_T^\ast ||$ and $|| \tilde \theta_2 - \theta_T^\ast || \le || \theta - \theta_T^\ast ||$, we have shown that for all $T$ sufficiently large enough, there exists a constant $c_1 > 0$ such that
	\begin{align}
	\left| \left| \Delta_T \big(\tilde \theta_1, \tilde \theta_2 \big) - \Lambda_T(\theta^\ast_T)  \right| \right| \le c_1 || \theta - \theta_T^\ast ||.
	\end{align}
	Furthermore, as the matrix $\Lambda_T(\theta^\ast_T)$ has Eigenvalues bounded from below (for $T$ large enough) by assumption, there exists a constant $c_2 > 0$, such that
	\begin{align}
	\left| \left| \Lambda_T(\theta^\ast_T)  \cdot (\theta - \theta_T^\ast) \right| \right| \ge c_2 ||\theta - \theta_T^\ast ||.
	\end{align}
	Thus, we choose $d_0 > 0$ small enough such that $d_0 < \frac{c_2}{2 c_1}$.
	Then $|| \theta - \theta_T^\ast|| \le d_0 < \frac{c_2}{2 c_1}$ and thus, $2c_1 ||\theta - \theta_T^\ast||^2 \le c_2 ||\theta - \theta_T^\ast ||$.
	Consequently, $\left| \left|\big( \Delta_T \big(\tilde \theta_1, \tilde \theta_2 \big)  - \Lambda_T(\theta^\ast_T)  \big) \cdot (\theta - \theta_T^\ast) \right| \right| \le c_1 ||\theta - \theta_T^\ast||^2 \le c_2/2 ||\theta - \theta_T^\ast||$ and thus 
	\begin{align}
	\big| \big| \Psi_{T}^0(\theta) \big| \big| 
	&= \big| \big| \Delta_T \big(\tilde \theta_1,\tilde \theta_2 \big) \cdot (\theta - \theta_T^\ast)\big| \big| \\
	&= \left| \left| \Lambda_T(\theta^\ast_T)  \cdot (\theta - \theta_T^\ast) + \big( \Delta_T \big(\tilde \theta_1,\tilde \theta_2 \big)  - \Lambda_T(\theta^\ast_T)  \big) \cdot (\theta - \theta_T^\ast )\right| \right| \\
	&\ge \Big| \left| \left| \Lambda_T(\theta^\ast_T)  \cdot (\theta - \theta_T^\ast) \right| \right| - \left| \left| \big( \Delta_T \big(\tilde \theta_1,\tilde \theta_2 \big)  - \Lambda_T(\theta^\ast_T)  \big) \cdot (\theta - \theta_T^\ast ) \right| \right| \Big| \\
	&\ge \frac{c_2}{2} ||\theta - \theta_T^\ast||,
	\end{align}
	by applying the mean value expansion and the inverse triangular inequality.
\end{proof}

\begin{lemma}
	\label{lemma:AsymptoticNormalityPsiESReg}
	Given Assumption \ref{ass:Assumption1ESReg}, it holds that
	\begin{align}
	\Sigma_T^{-1/2}(\theta_T^\ast) \,\sqrt{T} \, \Psi_T(\theta^\ast_T) \stackrel{d}{\to} \mathcal{N}(0,I_{2k}).
	\end{align}
\end{lemma}

\begin{proof}
	We show this multivariate result by applying the Cramér–Wold theorem, i.e. by showing that the conditions for the univariate CLT for $\alpha$-mixing sequences given in Theorem 5.20 in \cite{White2001}, p.130 hold for all linear combinations $u^\top \psi \big(Y_t, X_t, \theta_T^\ast \big)$ for all $u \in \mathbb{R}^k$ such that $||u|| = 1$.	
	By Theorem 3.49 in \cite{White2001} p.50, we get that the sequences $\psi \big(Y_t, X_t, \theta_T^\ast \big)$ and $u^\top \psi \big(Y_t, X_t, \theta_T^\ast \big)$ are strong mixing of size $-r/(r-2)$ for some $r > 2$.	
	Furthermore, for all $t \in \mathbb{N}$, it holds that 
	\begin{align*}
	&\mathbb{E} \left[ \big| u^\top \psi \big(Y_t, X_t, \theta_T^\ast \big) \big) \big|^{r} \right] 
	\le \mathbb{E} \left[ \big|\big| \psi \big(Y_t, X_t, \theta_T^\ast \big)\big) \big| \big|^{r} \right] \\
	\le \, &4^{r-1} \left\{
	\max \left( \frac{1-\tau}{\tau},1\right)^r \mathbb{E} \left[   \left| \left| \frac{V_t}{W_t^\top \gamma_T^\ast} \right| \right|^r \right]
	+ \mathbb{E} \left[   \left| \left| \frac{W_t W_t^\top \gamma_T^\ast}{(W_t^\top \gamma_T^\ast)^2} \right| \right|^r \right] \right. \\
	&\qquad \qquad + \left. \left( 1 + \frac{1}{\tau}\right)^r  \mathbb{E} \left[   \left| \left| \frac{W_t V_t^\top \beta_T^\ast}{(W_t^\top \gamma_T^\ast)^2} \right| \right|^r \right]
	+ \mathbb{E} \left[   \left| \left| \frac{W_t Y_t}{\tau (W_t^\top \gamma_T^\ast)^2} \right| \right|^r \right]
	\right\} \\
	\le \, &4^{r-1} \left\{
	\max \left( \frac{1-\tau}{\tau},1\right)^r \frac{1}{K^{r}} \mathbb{E} \left[   \left| \left| V_t \right| \right|^r \right]
	+ \frac{1}{K^{r}} \mathbb{E} \left[ \left| \left| W_t \right| \right|^r \right] \right. \\
	&\qquad \qquad + \left.  \frac{1}{K^{2r}} \left( 1 + \frac{1}{\tau}\right)^r  \mathbb{E} \left[   \left| \left| W_t V_t^\top \right| \right|^r \right]
	+ \frac{1}{\tau K^{2r}} \mathbb{E} \left[   \left| \left| W_t Y_t \right| \right|^r \right]
	\right\} < \infty,
	\end{align*}
	by applying Jensen's inequality and by the moment conditions \ref{cond:MomentConditionsESReg} in Assumption \ref{ass:Assumption1ESReg}, where $r > 2$ (from condition \ref{cond:StrongMixingESReg}).
	As the sequence $\psi \big(Y_t, X_t, \theta_T^\ast \big)$ is uncorrelated by condition \ref{cond:UniqueMinimum} in Assumption \ref{ass:Assumption1ESReg}, we get that
	%
	%
	%
	for all $T \ge 1$,
	\begin{align}
	\operatorname{Var} \left( \frac{1}{\sqrt{T}} \sum_{t=1}^T \psi \big(Y_t, X_t, \theta_T^\ast \big) \right)
	= \frac{1}{T} \sum_{t=1}^T \mathbb{E} \left[ \psi \big(Y_t, X_t, \theta_T^\ast)\cdot \psi \big(Y_t, X_t, \theta_T^\ast)^\top \right] = \Sigma_T(\theta_T^\ast).
	\end{align}
	As $\Sigma_T(\theta_T^\ast)$ is real and symmetric and positive definite, it can be diagonalized with a real orthogonal matrix $S$, i.e. $S^\top \Sigma_T(\theta_T^\ast) S = D_T$, where $D_T$ is a diagonal matrix containing the Eigenvalues of $\Sigma_T(\theta_T^\ast)$, denoted by  $\{ \lambda_{1,T},\dots,\lambda_{k,T}\}$.
	Consequently, for any $u \in \mathbb{R}^k$,
	\begin{align}
	\operatorname{Var} \left( \frac{1}{\sqrt{T}} \sum_{t=1}^T u^\top \psi  \big(Y_t, X_t, \theta_T^\ast \big)  \right) = u^\top \Sigma_T( \theta_T^\ast ) u 
	= u^\top S^\top D_T S u 
	=  v^\top D_T v 
	> \min_{i=1,\dots,k} \lambda_{i,T},
	\end{align}
	where $v = S u$, i.e. $||v|| = 1$ as $S$ is orthogonal and where the Eigenvalues $\{ \lambda_{1,T},\dots,\lambda_{k,T}\}$ are bounded away from zero for $T$ sufficiently large.
	Thus, we can apply Theorem 5.20 in \cite{White2001} p. 130 for asymptotic normality of the sequences $u^\top \psi \big(Y_t,X_t, \theta_T^\ast \big)$ for all $u \in \mathbb{R}^k$ such that $||u|| = 1$. Applying the Cramér-Wold theorem concludes the proof.
\end{proof}

\section{Existing Backtests}
\label{sec:existing_backtests}

Over the past two decades and especially driven by the recent transition from VaR to ES in the Basel regulatory framework \citep{Basel2016, Basel2017}, a large literature on backtesting the ES has emerged.
These backtests are usually introduced with financial regulators in mind who need to verify the risk forecasts they receive from the financial institutions.
To be applicable by the regulatory authorities, a backtest for the risk measure ES thus follows Definition \ref{def::ProperBacktest} and only requires the observed return series and the ES forecasts as input variables.
However, many of the proposed backtests for the ES fail to have this property.
In particular, several tests require the whole return distribution (or equivalently the cumulative violation process $\int_{0}^{\tau} \mathds{1}_{\{Y_t \leq \hat v_t(p) \}}\, \mathrm{d} p$) \citep{Kerkhof2004, Wong2008, Graham2014, Acerbi2014, Du2017, Loeser2018, CostanzinoCurran2018}, multiple quntile levels \citep{Emmer2015, Costanzino2015, Kratz2018, CouperierLeymarie2019}, the VaR and the volatility \citep{McNeil2000,Nolde2017,Righi2013,Righi2015}, or the VaR \citep{McNeil2000,Nolde2017} in addition to the ES forecasts.
However, this information is not reported by the financial institutions and therefore, most of these tests can not be used by the regulators \citep{Aramonte2011,Basel2017}.

Furthermore, when more information than solely the ES forecasts is used for backtesting, a rejection of the null hypothesis does not necessarily imply that the ES forecasts are wrong.
More precisely, a rejection of the null implies that \textit{some} component of the input parameters is incorrect \citep[cf.][]{Nolde2017}.
A related concern is raised by \citet{Aramonte2011}, who note that financial institutions could be tempted to submit forecasts of this additional information chosen such that the tests have particularly low power, so that correctness of their internal model (and their issued ES forecasts) is not doubted.

Strictly following Definition \ref{def::ProperBacktest}, we would have to distinguish between backtests for the ES and joint backtests for the pair VaR and ES.
However, as the ES is strongly intertwined with the VaR (through its definition and through the joint elicitability), sensible forecasts for the ES are based on correctly specified VaR forecasts.
Consequently, it is reasonable to backtest both quantities jointly and thus, we compare the performance of our ESR backtests to existing joint VaR and ES backtests in the literature. 
In the subsequent two sections, we describe the exceedance residual (ER) backtests of \citet{McNeil2000} and the conditional calibration (CC) backtests of \citet{Nolde2017} in detail, since both have versions that only require VaR forecasts in addition to the ES.

\subsection{Testing the Exceedance Residuals}
\label{sec:er_test}

One of the first and still most frequently used tests for the ES is the exceedance residual (ER) backtest of \citet{McNeil2000}.
This approach is based on the ES-specified residuals that exceed the VaR, $er_t = \big( Y_t -  \hat e_t \big) \mathds{1}_{\{Y_t \le \hat v_t\}}$, which form a martingale difference sequence given that $\hat v_t$ and $\hat e_t$ are the true quantile and ES conditional on the information $\mathcal{F}_{t-1}$.
\citet{McNeil2000} further consider a second version that uses exceedance residuals standardized by a given volatility forecast, i.e. $er_t / \hat \sigma_t$.

This backtest tests whether the expected value of the (raw or standardized) ER, $\mu = \mathbb{E} [ er_t]$, is zero using the estimate $\hat{\mu} = 1 / (\sum_{t=1}^T \mathds{1}_{\{Y_t \le \hat v_t\}}) \sum_{t=1}^T er_t$ in conjunction with a bootstrap hypothesis test \citep[see][p.~224]{Efron1994}.
In the original paper, \citet{McNeil2000} propose to test $\mu$ against the one-sided alternative that $\mu$ is negative, i.e. that the issued ES forecasts are too risky (too small in absolute value).
However, in this paper we discuss both, tests based on one-sided and two-sided hypotheses, so that in addition to the original proposal, we also include a two-sided test,
\begin{align}
\begin{split}
\mathbb{H}_0^{2s}:   \mu = 0 \qquad     & \text{against} \qquad \mathbb{H}_1^{2s}:  \mu \neq 0, \quad \text{and} \\
\mathbb{H}_0^{1s}:   \mu  \geq 0 \qquad & \text{against} \qquad \mathbb{H}_1^{1s}:  \mu  < 0.
\end{split}    
\end{align}	
By Definition \ref{def::ProperBacktest}, the test using the standardized ER is in fact a joint backtest for the triple VaR, ES and volatility, whereas the test using the raw ER  is a joint backtest for the pair VaR and ES.
In light of the discussion above, the test using the raw ER is therefore preferred.
Nevertheless, in the simulation studies and the empirical application we apply both approaches and find that they perform alike.

Even though the intercept ESR test introduced in \Cref{sec:OneSidedTest} and the ER backtest appear to be similar, there is a subtle but crucial difference between the two test statistics.
For the intercept ESR test, we compute the empirical ES of $Y_t - \hat e_t$, i.e. the average of $Y_t - \hat e_t$ given that $Y_t - \hat e_t$ is smaller than its \textit{empirical $\tau$-quantile}.
In contrast, the ER backtest computes the average of $Y_t - \hat e_t$, given that $Y_t$ is smaller than the \textit{respective forecast for its $\tau$-quantile $\hat v_t$}.
This difference seems marginal, but it has severe consequences for the theoretical and empirical properties of the tests.

As we can write $\hat{\mu} = 1/\tilde T \sum_{t=1}^T Y_t \mathds{1}_{\{Y_t \le \hat v_t\}} - 1/\tilde T \sum_{t=1}^T \hat e_t \mathds{1}_{\{Y_t \le \hat v_t\}}$, where $\tilde T = \sum_{t=1}^T \mathds{1}_{\{Y_t \le \hat v_t\}}$, the ER backtest in fact compares the empirical average of $Y_t$ truncated at $\hat v_t$ to the average ES forecast $\hat e_t$, whenever there is a VaR violation.
Thus, this backtest rejects whenever the distance/relation between the VaR and ES-forecasts is incorrect.
However, simultaneous misspecifications of both forecasts, such as e.g. generated by misspecification of the volatility process in location scale models cannot be detected.
In the same spirit, the ER backtest cannot distinguish between correct forecasts for the VaR and ES at level $\tau$ and (correct) forecasts for a misspecified probability level $\tilde{\tau} \neq \tau$, as the given level $\tau$ does not influence the ER test statistic at all.
In contrast, by computing the empirical $\tau$-quantile of $Y_t - \hat e_t$ (instead of using the forecast $\hat v_t$), the intercept ESR test does not suffer from these shortcomings as can be observed in the simulation results in Section \ref{sec::ContinuousMisspecification}.

\subsection{Conditional Calibration Backtests}
\label{sec:cc_test}

\citet{Nolde2017} introduce the concept of conditional calibration (CC) based on strict identification functions (also known as moment conditions or estimating equations) of the respective functional and show that many classical backtests for risk measures can be unified using this concept.
For the pair VaR and ES at level $\tau \in (0,1)$, they choose the strict identification function
\begin{align}
V(Y,\, v,\, e) = 
\begin{pmatrix}
\tau - \mathds{1}_{\left\{Y \leq v\right\}}\\
e - v + \mathds{1}_{\left\{Y \leq v\right\}}(v - Y) / \tau
\end{pmatrix},
\end{align}
whose expectation is zero if and only if $v$ and $e$ equal the true VaR and ES of the random variable $Y$ respectively.
The CC backtest for forecasts for the VaR, $\hat v_t$ and for the ES, $\hat e_t$ is based on the hypotheses
\begin{align}
\begin{split}
\mathbb{H}_0^{2s}:  \mathbb{E} \big[ V(Y_t, \, \hat v_t,\, \hat e_t ) \mid \mathcal{F}_{t-1} \big] = 0 \qquad     & \text{against} \qquad \mathbb{E} \big[ V(Y_t,\, \hat v_t,\, \hat e_t ) \mid \mathcal{F}_{t-1} \big] \neq 0, \quad \text{and} \\
\mathbb{H}_0^{1s}:  \mathbb{E} \big[ V(Y_t, \, \hat v_t, \, \hat e_t ) \mid \mathcal{F}_{t-1} \big] \geq 0 \qquad & \text{against} \qquad \mathbb{E} \big[ V(Y_t, \, \hat v_t, \, \hat e_t ) \mid \mathcal{F}_{t-1} \big] < 0,
\end{split}    
\end{align}	
component-wise and almost surely for all $t= 1,\ldots,T$.
This is equivalent to testing $\mathbb{E} \big[ h_t^\top V(Y_t, \hat v_t, \hat e_t ) \big] = 0$ for all $\mathcal{F}_{t-1}$ measurable $\mathbb{R}^2$-valued functions $h_t$.
As this is infeasible, \citet{Nolde2017} propose to use an $\mathcal{F}_{t-1}$-measurable sequence of $q \times 2$-matrices of test functions $\boldsymbol{h}_t$ for some $q \in \mathbb{N}$ and to use the Wald-type test statistic
\begin{align}
T_{\text{CC}} = T \left(\frac{1}{T}\sum_{t=1}^T \boldsymbol{h}_t V \left(Y_t, \hat v_t,\hat e_t \right) \right)^\top \widehat{\Delta}_T^{-1} \left(\frac{1}{T}\sum_{t=1}^T \boldsymbol{h}_t V \left(Y_t, \hat v_t,\hat e_t \right) \right),
\end{align}
where $
\widehat{\Delta}_T = \frac{1}{T} \sum_{t=1}^{T} 
\big( \boldsymbol{h}_t V \left(Y_t, \hat v_t,\hat e_t \right) \big)
\big( \boldsymbol{h}_t V \left(Y_t, \hat v_t,\hat e_t \right) \big)^\top
$
is a consistent estimator of the covariance of the $q$-dimensional vector $\boldsymbol{h}_t V \left(Y_t, \hat v_t,\hat e_t \right)$.
Under $\mathbb{H}_0$, the test statistic asymptotically follows a $\chi^2_q$ distribution with $q$ degrees of freedom.

\citet{Nolde2017} propose two versions of this test, where the first uses no information besides the risk forecasts (termed \textit{simple CC test}), and where the second additionally requires volatility forecasts (termed \textit{general CC} test).
For the simple CC test, the test function is the identity matrix, $\boldsymbol{h}_t = I_2,$ for both, the one- and two-sided hypotheses.
For the general CC test, they propose to choose
\begin{align}
\boldsymbol{h}_t  = \hat \sigma_t \big(\left(\hat e_t - \hat v_t \right) / \tau,\, 1 \big) \quad \text{and} \quad \boldsymbol{h}_t  = \begin{pmatrix}
1 & |\hat v_t| & 0 & 0 \\
0 & 0 & 1 & \hat \sigma_t^{-1}
\end{pmatrix}^\top,
\end{align}
for the two-sided and for the one-sided test, respectively, where $\hat \sigma_t$ is a forecast for the volatility.
As with the standardized ER test, the general CC test is strictly speaking a backtest for the triple VaR, ES, and volatility, but we nevertheless include both versions in our empirical comparisons.
We provide implementations of the two ESR backtests proposed in this paper, both ER backtests of \citet{McNeil2000} and both CC backtests of \citet{Nolde2017} in the R package \texttt{esback} \citep{Bayer2019a}.

\newpage

\section{Additional Material}
\label{sec:AdditionalMaterial}

\begin{table}[h]
	\footnotesize
	\centering
	\caption{Empirical sizes for the first simulation study.}
	\label{tab:mc1_size1Percent}
	\begin{tabularx}{\linewidth}{XX *{12}{r}}
		\toprule
		\theadl{DGP} & \theadl{Sample \\ Size} & \thead{Str. \\ ESR} & \thead{Aux. \\ ESR} & \thead{Int. \\ ESR} & \thead{Str. \\ ESR} & \thead{Aux. \\ ESR} & \thead{Int. \\ ESR}  & \thead{Gen. \\ CC} & \thead{Sim. \\ CC} & \thead{Std. \\ ER} & \thead{ER} \\
		\cmidrule(lr){3-5} \cmidrule(lr){6-8} \cmidrule(lr){9-12}
		& & \multicolumn{3}{c}{Misspec Covariance} & \multicolumn{3}{c}{Classical Covariance} \\	
		\midrule
            &   250 & 0.05 & 0.05 & 0.09 & 0.16 & 0.16 & 0.10 & 0.01 & 0.22 & 0.04 & 0.05 \\
            &   500 & 0.03 & 0.03 & 0.05 & 0.08 & 0.08 & 0.06 & 0.03 & 0.13 & 0.01 & 0.01 \\
 EGARCH-STD &  1000 & 0.02 & 0.02 & 0.03 & 0.05 & 0.05 & 0.03 & 0.04 & 0.08 & 0.01 & 0.02 \\
            &  2500 & 0.01 & 0.01 & 0.01 & 0.02 & 0.02 & 0.01 & 0.02 & 0.04 & 0.01 & 0.01 \\
            &  5000 & 0.01 & 0.01 & 0.01 & 0.02 & 0.01 & 0.01 & 0.02 & 0.03 & 0.01 & 0.01 \\
		\midrule
         &   250 & 0.05 & 0.06 & 0.08 & 0.17 & 0.17 & 0.10 & 0.01 & 0.21 & 0.04 & 0.05 \\
         &   500 & 0.04 & 0.04 & 0.06 & 0.09 & 0.09 & 0.06 & 0.02 & 0.13 & 0.01 & 0.01 \\
 GAS-STD &  1000 & 0.02 & 0.02 & 0.03 & 0.06 & 0.06 & 0.03 & 0.03 & 0.08 & 0.02 & 0.02 \\
         &  2500 & 0.01 & 0.02 & 0.01 & 0.03 & 0.03 & 0.01 & 0.02 & 0.05 & 0.01 & 0.01 \\
         &  5000 & 0.01 & 0.01 & 0.01 & 0.02 & 0.02 & 0.01 & 0.02 & 0.03 & 0.01 & 0.01 \\
		\midrule
          &   250 & 0.05 & 0.05 & 0.08 & 0.16 & 0.16 & 0.09 & 0.01 & 0.20 & 0.04 & 0.04 \\
          &   500 & 0.03 & 0.03 & 0.05 & 0.08 & 0.08 & 0.06 & 0.02 & 0.12 & 0.01 & 0.01 \\
 GAS-SSTD &  1000 & 0.02 & 0.02 & 0.03 & 0.04 & 0.04 & 0.03 & 0.03 & 0.07 & 0.02 & 0.01 \\
          &  2500 & 0.01 & 0.01 & 0.01 & 0.02 & 0.02 & 0.01 & 0.02 & 0.04 & 0.01 & 0.01 \\
          &  5000 & 0.01 & 0.01 & 0.01 & 0.01 & 0.01 & 0.01 & 0.02 & 0.03 & 0.02 & 0.01 \\
		\midrule
                      &   250 & 0.02 & 0.02 & 0.06 & 0.10 & 0.10 & 0.07 & 0.01 & 0.17 & 0.04 & 0.04 \\
                      &   500 & 0.02 & 0.02 & 0.04 & 0.05 & 0.05 & 0.05 & 0.02 & 0.09 & 0.00 & 0.00 \\
 AR-GARCH, $\phi=0.0$ &  1000 & 0.01 & 0.01 & 0.03 & 0.04 & 0.04 & 0.03 & 0.02 & 0.05 & 0.00 & 0.01 \\
                      &  2500 & 0.01 & 0.01 & 0.02 & 0.02 & 0.02 & 0.02 & 0.02 & 0.03 & 0.01 & 0.01 \\
                      &  5000 & 0.01 & 0.01 & 0.01 & 0.02 & 0.02 & 0.01 & 0.01 & 0.02 & 0.01 & 0.01 \\
		\midrule
                      &   250 & 0.02 & 0.02 & 0.06 & 0.10 & 0.10 & 0.07 & 0.01 & 0.17 & 0.04 & 0.04 \\
                      &   500 & 0.02 & 0.02 & 0.04 & 0.05 & 0.05 & 0.05 & 0.02 & 0.09 & 0.00 & 0.00 \\
 AR-GARCH, $\phi=0.1$ &  1000 & 0.01 & 0.01 & 0.03 & 0.03 & 0.04 & 0.03 & 0.02 & 0.05 & 0.00 & 0.01 \\
                      &  2500 & 0.01 & 0.01 & 0.02 & 0.02 & 0.02 & 0.02 & 0.02 & 0.03 & 0.01 & 0.01 \\
                      &  5000 & 0.01 & 0.01 & 0.01 & 0.01 & 0.02 & 0.01 & 0.01 & 0.02 & 0.01 & 0.01 \\
		\midrule
                      &   250 & 0.02 & 0.02 & 0.06 & 0.09 & 0.09 & 0.07 & 0.01 & 0.17 & 0.04 & 0.04 \\
                      &   500 & 0.02 & 0.02 & 0.04 & 0.06 & 0.06 & 0.04 & 0.02 & 0.09 & 0.00 & 0.00 \\
 AR-GARCH, $\phi=0.5$ &  1000 & 0.01 & 0.01 & 0.03 & 0.04 & 0.04 & 0.03 & 0.02 & 0.05 & 0.00 & 0.01 \\
                      &  2500 & 0.01 & 0.01 & 0.02 & 0.02 & 0.02 & 0.02 & 0.02 & 0.03 & 0.01 & 0.01 \\
                      &  5000 & 0.01 & 0.01 & 0.01 & 0.02 & 0.02 & 0.01 & 0.01 & 0.02 & 0.01 & 0.01 \\
		\bottomrule
		\addlinespace
		\multicolumn{12}{p{.97\linewidth}}{\textit{Notes:} The table reports the empirical sizes of the backtests for the different DGPs decribed in Section \ref{sec::TraditionalSizePower} and for a nominal test size of $1\%$.
			The number of Monte-Carlo repetitions is 10,000 and the probability level for the risk measures is $\tau=2.5\%$. 
			ESR refers to the three backtests introduced in this paper and we consider versions with covariance estimation with and without model misspecification.
			CC refers to the conditional calibration tests of \citet{Nolde2017}, and ER to the exceedance residuals tests of \citet{McNeil2000}.}
	\end{tabularx}
\end{table}

\begin{table}
	\footnotesize
	\centering
	\caption{Empirical sizes for the first simulation study.}
	\label{tab:mc1_size10Percent}
	\begin{tabularx}{\linewidth}{XX *{12}{r}}
		\toprule
		\theadl{DGP} & \theadl{Sample \\ Size} & \thead{Str. \\ ESR} & \thead{Aux. \\ ESR} & \thead{Int. \\ ESR} & \thead{Str. \\ ESR} & \thead{Aux. \\ ESR} & \thead{Int. \\ ESR}  & \thead{Gen. \\ CC} & \thead{Sim. \\ CC} & \thead{Std. \\ ER} & \thead{ER} \\
		\cmidrule(lr){3-5} \cmidrule(lr){6-8} \cmidrule(lr){9-12}
		& & \multicolumn{3}{c}{Misspec Covariance} & \multicolumn{3}{c}{Classical Covariance} \\	
		\midrule
            &   250 & 0.12 & 0.12 & 0.19 & 0.31 & 0.31 & 0.20 & 0.16 & 0.33 & 0.12 & 0.14 \\
            &   500 & 0.09 & 0.09 & 0.15 & 0.21 & 0.21 & 0.16 & 0.16 & 0.25 & 0.10 & 0.12 \\
 EGARCH-STD &  1000 & 0.08 & 0.08 & 0.13 & 0.16 & 0.16 & 0.13 & 0.14 & 0.19 & 0.10 & 0.13 \\
            &  2500 & 0.08 & 0.08 & 0.11 & 0.11 & 0.12 & 0.11 & 0.12 & 0.14 & 0.10 & 0.11 \\
            &  5000 & 0.08 & 0.08 & 0.15 & 0.10 & 0.10 & 0.15 & 0.11 & 0.13 & 0.10 & 0.11 \\
		\midrule
         &   250 & 0.14 & 0.14 & 0.18 & 0.32 & 0.32 & 0.20 & 0.15 & 0.32 & 0.13 & 0.13 \\
         &   500 & 0.11 & 0.11 & 0.14 & 0.22 & 0.22 & 0.15 & 0.15 & 0.25 & 0.12 & 0.12 \\
 GAS-STD &  1000 & 0.09 & 0.09 & 0.12 & 0.16 & 0.16 & 0.12 & 0.14 & 0.19 & 0.12 & 0.12 \\
         &  2500 & 0.09 & 0.09 & 0.12 & 0.13 & 0.12 & 0.12 & 0.12 & 0.15 & 0.11 & 0.12 \\
         &  5000 & 0.09 & 0.09 & 0.17 & 0.11 & 0.11 & 0.17 & 0.11 & 0.13 & 0.11 & 0.11 \\
		\midrule
          &   250 & 0.12 & 0.12 & 0.17 & 0.31 & 0.31 & 0.19 & 0.15 & 0.30 & 0.14 & 0.13 \\
          &   500 & 0.09 & 0.09 & 0.14 & 0.20 & 0.20 & 0.15 & 0.15 & 0.23 & 0.12 & 0.11 \\
 GAS-SSTD &  1000 & 0.08 & 0.08 & 0.12 & 0.15 & 0.15 & 0.12 & 0.13 & 0.18 & 0.12 & 0.11 \\
          &  2500 & 0.07 & 0.07 & 0.11 & 0.11 & 0.11 & 0.11 & 0.13 & 0.14 & 0.12 & 0.10 \\
          &  5000 & 0.08 & 0.08 & 0.13 & 0.10 & 0.10 & 0.13 & 0.12 & 0.12 & 0.11 & 0.10 \\
		\midrule
                      &   250 & 0.07 & 0.07 & 0.16 & 0.24 & 0.24 & 0.18 & 0.13 & 0.26 & 0.11 & 0.11 \\
                      &   500 & 0.07 & 0.07 & 0.14 & 0.18 & 0.18 & 0.15 & 0.13 & 0.19 & 0.08 & 0.09 \\
 AR-GARCH, $\phi=0.0$ &  1000 & 0.07 & 0.07 & 0.12 & 0.14 & 0.14 & 0.12 & 0.12 & 0.16 & 0.09 & 0.09 \\
                      &  2500 & 0.07 & 0.07 & 0.11 & 0.11 & 0.11 & 0.11 & 0.11 & 0.12 & 0.10 & 0.10 \\
                      &  5000 & 0.08 & 0.08 & 0.11 & 0.11 & 0.11 & 0.11 & 0.11 & 0.11 & 0.10 & 0.10 \\
		\midrule
                      &   250 & 0.07 & 0.07 & 0.16 & 0.24 & 0.24 & 0.18 & 0.13 & 0.26 & 0.11 & 0.11 \\
                      &   500 & 0.07 & 0.07 & 0.14 & 0.18 & 0.18 & 0.15 & 0.13 & 0.19 & 0.08 & 0.09 \\
 AR-GARCH, $\phi=0.1$ &  1000 & 0.06 & 0.07 & 0.12 & 0.14 & 0.14 & 0.12 & 0.12 & 0.16 & 0.09 & 0.09 \\
                      &  2500 & 0.07 & 0.07 & 0.11 & 0.11 & 0.11 & 0.11 & 0.11 & 0.12 & 0.10 & 0.10 \\
                      &  5000 & 0.08 & 0.07 & 0.11 & 0.11 & 0.11 & 0.11 & 0.11 & 0.11 & 0.10 & 0.10 \\
		\midrule
                      &   250 & 0.06 & 0.06 & 0.16 & 0.23 & 0.23 & 0.18 & 0.13 & 0.26 & 0.11 & 0.11 \\
                      &   500 & 0.06 & 0.06 & 0.14 & 0.18 & 0.18 & 0.15 & 0.13 & 0.19 & 0.08 & 0.09 \\
 AR-GARCH, $\phi=0.5$ &  1000 & 0.06 & 0.06 & 0.12 & 0.14 & 0.14 & 0.12 & 0.12 & 0.16 & 0.09 & 0.09 \\
                      &  2500 & 0.07 & 0.07 & 0.11 & 0.12 & 0.12 & 0.11 & 0.11 & 0.12 & 0.10 & 0.10 \\
                      &  5000 & 0.08 & 0.08 & 0.11 & 0.12 & 0.11 & 0.11 & 0.11 & 0.11 & 0.10 & 0.10 \\
		\bottomrule
		\addlinespace
		\multicolumn{12}{p{.97\linewidth}}{\textit{Notes:} The table reports the empirical sizes of the backtests for the different DGPs decribed in Section \ref{sec::TraditionalSizePower} and for a nominal test size of $10\%$.
			The number of Monte-Carlo repetitions is 10,000 and the probability level for the risk measures is $\tau=2.5\%$. 
			ESR refers to the three backtests introduced in this paper and we consider versions with covariance estimation with and without model misspecification.
			CC refers to the conditional calibration tests of \citet{Nolde2017}, and ER to the exceedance residuals tests of \citet{McNeil2000}.}
	\end{tabularx}
\end{table}

\begin{figure}[h]
	\centering
	\begin{subfigure}{.49\linewidth}
		\caption{Changing the reaction to the squared returns}
		\label{fig:mc2_example_series_1}
		\includegraphics{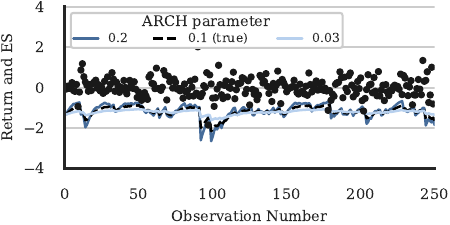}
	\end{subfigure}
	\hfill
	\begin{subfigure}{.49\linewidth}
		\caption{Changing the unconditional variance}
		\label{fig:mc2_example_series_2}
		\includegraphics{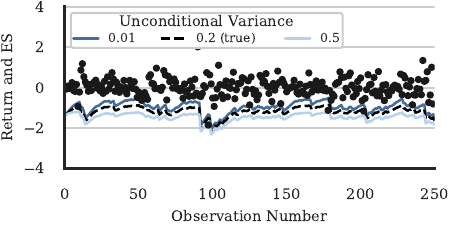}
	\end{subfigure}
	\\[\baselineskip]
	\begin{subfigure}{.49\linewidth}
		\caption{Changing the persistence}    
		\label{fig:mc2_example_series_3}
		\includegraphics{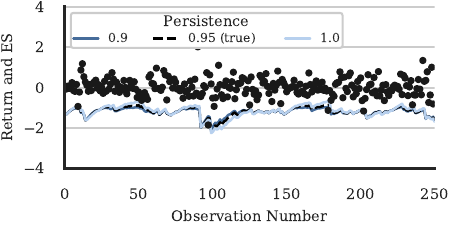}
	\end{subfigure}	
	\\[\baselineskip]
	\begin{subfigure}{.49\linewidth}
		\caption{Changing the degrees of freedom}
		\label{fig:mc2_example_series_4}
		\includegraphics{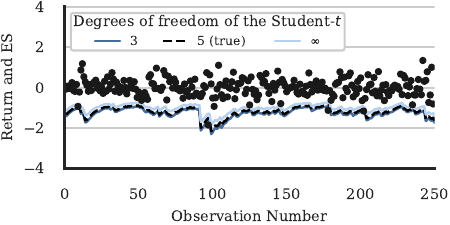}
	\end{subfigure}	
	\hfill
	\begin{subfigure}{.49\linewidth}
		\caption{Changing the probability level}
		\label{fig:mc2_example_series_5}
		\includegraphics{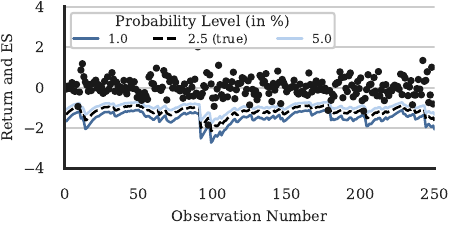}
	\end{subfigure}	
	\caption[Simulated Return Series under Misspecifications.]{These plots show exemplary simulated return series with 250 observations for the DGP given in (\ref{eq:mc2_model}) and for the five parameter misspecifications illustrated in the points (a) - (e) in Section \ref{sec::ContinuousMisspecification}. 	In each of the subfigures, the black dashed line corresponds to the true model parameters.}
	\label{fig:mc2_example_series}
\end{figure}

\end{appendices}

\FloatBarrier

\addcontentsline{toc}{section}{References}
\bibliographystyle{apalike}	
\bibliography{mybib_ESback}

\end{document}